%% file: manuscript-numapde-preprint.tex
\pdfoutput=1

\makeatletter
\IfFileExists{./numapde-latex/numapde-packages.sty}{\providecommand*{\input@path}{}\edef\input@path{{./numapde-latex/}\input@path}}{}
\makeatother

\documentclass[english]{numapde-preprint}

\usepackage{numapde-semantic}
\usepackage{numapde-style}
\usepackage{numapde-local}

\IfFileExists{./numapde-bibliography/numapde.bib}{\addbibresource{./numapde-bibliography/numapde.bib}}{\addbibresource{numapde.bib}}

\hypersetup{
	pdftitle={Planning of Measurement Series for Thermodynamic Properties based on Optimal Experimental Design},
	pdfauthor={Ophelia Frotscher, Roland Herzog, Markus Richter},
	pdfkeywords={optimal experimental design, correlation, density, ethylene gylcol, measurement}
}

\title{Planning of Measurement Series for Thermodynamic Properties based on Optimal Experimental Design}
\subtitle{}
\shorttitle{Measurement Series for Thermodynamic Properties based on OED}

\author{Ophelia Frotscher\thanks{Technische Universität Chemnitz, Faculty of Mechanical Engineering, Professorship Applied Thermodynamics, 09107 Chemnitz, Germany (\email{ophelia.frotscher@mb.tu-chemnitz.de}, \url{https://www.tu-chemnitz.de/mb/TechnThDyn/personal.php}, \orcid{0000-0002-6915-1988}, \email{m.richter@mb.tu-chemnitz.de}, \url{https://www.tu-chemnitz.de/mb/TechnThDyn/personal.php}, \orcid{0000-0001-8120-5646}).}
\and
Roland Herzog\thanks{Technische Universität Chemnitz, Faculty of Mathematics, Professorship Numerical Mathematics (Partial Differential Equations), 09107 Chemnitz, Germany (\email{roland.herzog@mathematik.tu-chemnitz.de}, \url{https://www.tu-chemnitz.de/mathematik/part_dgl/people/herzog}, \orcid{0000-0003-2164-6575}).}
\and
Markus Richter\footnotemark[1]}
\shortauthor{O. Frotscher, R. Herzog and M. Richter}

\dedication{}

\IfFileExists{numapde-uncolorizeHyperrefIfChanges.sty}{\RequirePackage{numapde-uncolorizeHyperrefIfChanges}}{}

\begin{document}
\maketitle

\begin{abstract}
Decreasing the time required for accurate thermodynamic property measurements is extremely desirable for model development, that can respond to the needs of science and industry within a short time frame.
Here, we demonstrate the application of optimal experimental design to measurements of thermodynamic properties.
The technique is exemplified using the fitting of a Schilling-type equation from published \((p, \rho, T)\)-measurements of ethylene gylcol and propylene gylcol.
The analysis shows that a fixed-exponent fit using \((p,T)\)-measurements along the five most informative isotherms produces models of relative density errors comparable to those obtained using the data along all isotherms.
It is also argued that a calculation of optimal isotherms prior to the measurement series can further increase the precision at no additional experimental effort.\end{abstract}



\input{main.tex}

\printbibliography

\end{document}

%% file: main.tex

\section{Introduction}
\label{section:introduction}
\input{01_introduction.tex}

\section{Modeling Approach}
\label{section:classical_correlation_models}
\input{02_correlation_models.tex}

\section{Background on Optimal Experimental Design}
\label{section:oed}
\input{03_background_OED.tex}

\section{Numerical Results}
\label{section:numerical_results}
\input{04_numerical_results.tex}

\section{Conclusions}
\label{section:conclusion}
\input{05_conclusion.tex}

%% file: 01_introduction.tex
For scientific but particularly for industrial applications, it is absolutely valuable to provide equations of state (EOS) that accurately describe the thermodynamic behavior of the concerned fluid substances within a narrow time frame.
However, accurate thermodynamic models are usually based on reliable experimental data, thus, the development can require quite a long time since accurate measurements over wide temperature and pressure ranges are typically a time-consuming endeavor.
Here, we demonstrate the advantages of involving optimal experimental design (OED) in the process of planning measurement series for thermodynamic properties with the goal to deliberately reduce the experimental effort, thus, decreasing the overall time for model development.
Therefore, the issue with typical experimental design and the approach of OED will be explained using our density data of ethylene glycol published by Yang et al.\  \cite{YangSampsonFrotscherRichter:2020:1} and the density data of propylene glycol measured with the same instrument published by Sampson et al.\ \cite{SampsonYangXuRichter:2019:1}.
The application of OED is described using the measurements for ethylene glycol, which are shown in \cref{fig:expData}.
The \((p, \rho, T)\) behavior of ethylene glycol was investigated over the temperature range from \(T = \)~(\numrange{283.3}{393.1})~\si{\kelvin} at pressures from \(p =\)~(\numrange{4.8}{100.1})~\si{\mega\pascal} utilizing a high-pressure vibrating-tube densimeter; a total of \num{89} \((p, \rho, T)\) data points were studied with a combined expanded uncertainty (\(k=2\)) of \SI{1.57}{\kilo\gram\per\meter\cubed} (equivalent to a max.\ relative uncertainty of \SI[round-mode=figures,round-precision=3]{0.150823}{\%}).
To model the experimental data, in Yang et al.\ \cite{YangSampsonFrotscherRichter:2020:1} two empirical Schilling-type correlation equations were fitted \cite{SchillingKleinrahmWagner:2008:1}: one of the same form (same number of terms, same exponents) as for propylene glycol in the work of Sampson et al.\ \cite{SampsonYangXuRichter:2019:1} (fixed exponent model), and one using the "artificial intelligence powered" software tool Eureqa (Eureqa model) \cite{SchmidtLipson:2009:1,Eureqa:2015}.
The first approach results from the assumed thermodynamic similarity of both substances, and the second approach serves to investigate the applicability of symbolic regression and machine learning for the optimization of correlation equations.
\Cref{section:classical_correlation_models} of the present paper provides a small overview of different types of correlation models for liquid-phase densities to support the understanding of the choices made in this work. 
 
\begin{figure}
	\centering
	\input{matlab/tikz_expData.tex}
	\caption{Experimental densities (symbols) and interpolated densities (dashed lines) measured by Yang et al.\ \cite{YangSampsonFrotscherRichter:2020:1}. \textcolor{black}{\rotatebox[origin=c]{180}{\(\triangle\)}}, \(T\approx\)\,\SI{283.32}{\kelvin}; \textcolor{mycolor1}{\(+\)}, \(T\approx\)\,\SI{293.14}{\kelvin}; \textcolor{mycolor1}{\rotatebox[origin=c]{270}{\(\triangle\)}}, \(T\approx\)\,\SI{293.14}{\kelvin}; \textcolor{mycolor2}{o}, \(T\approx\)\,\SI{298.18}{\kelvin}; \textcolor{mycolor3}{\(*\)}, \(T\approx\)\,\SI{313.13}{\kelvin}; \textcolor{mycolor4}{x}, \(T\approx\)\,\SI{333.09}{\kelvin}; \textcolor{mycolor5}{\(\boxempty\)}, \(T\approx\)\,\SI{353.13}{\kelvin}; \textcolor{mycolor6}{\(\diamond\)}, \(T\approx\)\,\SI{373.22}{\kelvin}; \textcolor{mycolor7}{\(\triangle\)}, \(T\approx\)\,\SI{393.09}{\kelvin}}
	\label{fig:expData}
\end{figure}

In the present study, we illustrate how OED could have been used for the planning of the measurement series conducted by Yang et al. \cite{YangSampsonFrotscherRichter:2020:1} in order to reduce the amount of density measurements of ethylene glycol needed to reliably adjust the parameters of the fixed exponents model.
To validate our results regarding the benefits of OED, the calculations were also applied to measured densities of propylene glycol \cite{SampsonYangXuRichter:2019:1}.
In spite of its capabilities to select measurements with the highest information content, OED is hardly applied in the field of thermodynamic property research. We conjecture that the primary reasons for this are:

\begin{itemize}
	\item 
		a lack of knowledge about OED technologies,
	\item 
		the lack of software tools for easy application of OED,
	\item
		the requirement to specify the underlying model beforehand.
\end{itemize}

For a detailed overview of the use of OED in different areas of chemical and thermal engineering research, we refer the reader to Francheschini and Macchietto \cite{FranceschiniMacchietto:2008:1}.
With respect to thermodynamic property research, we point out the work of Bardow and colleagues \cite{RaschBueckerBardow:2009:1,DechambreWolffPaulsBardow:2014:1}, which deals with the experimental characterization of liquid-liquid equilibria utilizing OED.
In contrast to those investigations, the present work focuses on the temperature and pressure values for density measurements, with the aim to minimize the parameter variance of the correlation model.
The method described here can be used equivalently for measurements of other thermodynamic and transport properties (\eg, speed of sound, specific heat capacity, viscosity, etc.).

When changing the isotherm in a usual measurement series, establishing thermal equilibrium with a typical \((p, \rho, T)\) apparatus is rather time-consuming compared to setting new pressures along an isotherm. For this reason, we decided to use OED to select a subset of the most informative isotherms.
OED can be imagined as an in-situ tool within the following workflow:
\begin{enumeratelatin}
	\item
		choose an initial set of \((p, T)\) state points to be studied (\eg, two isotherms à five pressures),
	\item
		\label{item:conduct_measurements}
		conduct the measurements (\eg, density),
	\item
		fit the model (\eg, the Schilling-type model),
	\item
		answer the following questions:
		\begin{itemize}	
			\item 
				Can the model reproduce the measured data with sufficient accuracy (\eg, within the experimental uncertainty)?
			\item
				Can a significant amount of information be gained from additional measurements?
		\end{itemize}
	\item
		If the latter is not the case, the experiment is terminated.
		Otherwise, select the next most relevant isotherm using OED (\cref{section:oed}) and proceed with \cref{item:conduct_measurements}.
\end{enumeratelatin}

In this paper, we utilize OED to select a subset of isothermal experiments from a two given sets of measurements, which are already available.
We use the measurements selected to fit the parameters for new fixed and free exponent models for ethylene glycol and propylene glycol.
The results are compared to the existing models (\cref{section:numerical_results}).

%% file: matlab/tikz_expData.tex
%
%
\definecolor{mycolor1}{rgb}{0.49400,0.18400,0.55600}%
\definecolor{mycolor2}{rgb}{1.00000,0.00000,1.00000}%
\definecolor{mycolor3}{rgb}{0.00000,0.44700,0.74100}%
\definecolor{mycolor4}{rgb}{0.30100,0.74500,0.93300}%
\definecolor{mycolor5}{rgb}{0.46600,0.67400,0.18800}%
\definecolor{mycolor6}{rgb}{0.92900,0.69400,0.12500}%
\definecolor{mycolor7}{rgb}{0.63500,0.07800,0.18400}%
\begin{tikzpicture}

\begin{axis}[%
width=0.97\linewidth,
height=0.6\linewidth,
xmin=0,
xmax=105,
xlabel style={font=\color{white!15!black}},
xlabel={Pressure $p$ / \si{\mega\pascal}},
ymin=1030,
ymax=1160,
ylabel style={font=\color{white!15!black},at={(axis description cs:-0.02,0.45)}},
ylabel={Density $\rho_{\text{exp}}$ / \si{\kilo\gram\per\meter\cubed}},
axis background/.style={fill=white}
]
\addplot [color=black, dashed, line width=1.0pt, mark=triangle, mark options={solid, rotate=180, black}, forget plot]
  table[row sep=crcr]{%
5.28	1120.61\\
9.85	1122.37\\
15.02	1124.33\\
20.09	1126.22\\
30.12	1129.86\\	
49.89	1136.79\\
70.07	1143.48\\
89.83	1149.79\\
100.03	1152.9\\
};
\addplot [color=mycolor1, dashed, line width=1.0pt, mark=+, mark options={solid, mycolor1}, forget plot]
  table[row sep=crcr]{%
4.98	1113.6\\
10.1	1115.64\\
15.15	1117.63\\
19.96	1119.45\\
30.15	1123.25\\
49.88	1130.32\\
69.49	1136.99\\
89.98	1143.65\\
99.98	1146.77\\
};
\addplot [color=mycolor1, dashed, line width=1.0pt, mark=triangle, mark options={solid, rotate=270, mycolor1}, forget plot]
table[row sep=crcr]{%
5.04	1113.69\\
10.16	1115.7\\
15	1117.54\\
19.84	1119.4\\
29.94	1123.17\\
49.86	1130.31\\
69.97	1137.12\\
89.75	1143.54\\
100	1146.75\\
};
\addplot [color=mycolor2, dashed, line width=1.0pt, mark=o, mark options={solid, mycolor2}, forget plot]
  table[row sep=crcr]{%
5.08	1110.15\\
10.25	1112.2\\
15.33	1114.19\\
19.91	1115.97\\
30.18	1119.86\\
49.93	1127.05\\
70.05	1133.92\\
89.99	1140.47\\
99.94	1143.6\\
};
\addplot [color=mycolor3, dashed, line width=1.0pt, mark=asterisk, mark options={solid, mycolor3}, forget plot]
  table[row sep=crcr]{%
5.05	1099.85\\
10.03	1101.91\\
15.02	1103.96\\
20.05	1105.98\\
30.19	1109.97\\
49.84	1117.39\\
69.9	1124.56\\
89.94	1131.37\\
99.65	1134.53\\
};
\addplot [color=mycolor4, dashed, line width=1.0pt, mark=x, mark options={solid, mycolor4}, forget plot]
  table[row sep=crcr]{%
4.83	1085.85\\
10.03	1088.16\\
15.07	1090.35\\
20.07	1092.5\\
29.96	1096.63\\
50.01	1104.65\\
70.18	1112.21\\
90.19	1119.35\\
100.1	1122.76\\
};
\addplot [color=mycolor5, dashed, line width=1.0pt, mark=square, mark options={solid, mycolor5}, forget plot]
  table[row sep=crcr]{%
4.94	1071.71\\
9.87	1074.06\\
15.22	1076.53\\
19.97	1078.71\\
30.16	1083.24\\
49.89	1091.67\\
70.02	1099.59\\
89.98	1107.08\\
100.01	1110.71\\
};
\addplot [color=mycolor6, dashed, line width=1.0pt, mark=diamond, mark options={solid, mycolor6}, forget plot]
  table[row sep=crcr]{%
5.04	1056.86\\
9.99	1059.39\\
14.98	1061.79\\
19.91	1064.21\\
30.04	1069.04\\
49.96	1078.1\\
69.96	1086.5\\
90	1094.42\\
100.05	1098.23\\
};
\addplot [color=mycolor7, dashed, line width=1.0pt, mark=triangle, mark options={solid, mycolor7}, forget plot]
  table[row sep=crcr]{%
5	1040.95\\
9.84	1044.16\\
15.1	1047\\
19.95	1049.57\\
29.95	1054.69\\
49.93	1064.3\\
69.82	1073.17\\
90	1081.61\\
100.07	1085.56\\
};
\end{axis}
\end{tikzpicture}%

%% file: 02_correlation_models.tex
For simulations in process engineering, accurate and easily applicable models that describe the real fluid behavior are needed.
Considering the lack of knowledge about more complex compounds at the molecular level, empirical equations of state have become an established way of modeling thermodynamic behavior.
In industrial applications, the empirical Tait equation (1888) is still widely used to model densities of liquids \cite{GibsonLoeffler:1939:1}. 
However, a contemporary approach to model liquid-phase densities was developed by Schilling et al.\ \cite{SchillingKleinrahmWagner:2008:1}.
This is an equation in form of polynomial terms, which is based on the relationship to fundamental equations of state.

For the modeling of fundamental equations of state, which provide the possibility to derive all thermodynamic properties by differentiation of these equations, functional forms such as polynomial and exponential terms are involved \cite{Span:2000:1,TholBell:2020:1}.
In the past, within the scope of developing fundamental equations, Setzmann and Wagner (1989) \cite{SetzmannWagner:1989:1,Span:2000:1} established the modeling tool OPTIM, which combines a modified step-wise regression analysis based on a \eqq{\emph{bank of terms}} with elements of evolutionary optimization methods. 
Nowadays, this tool is mostly unused.
However, it was applied to model the density of liquid $n$-heptane, $n$-nonane, 2,4-dichlorotoluene and bromobenzene in the temperature range from \numrange{233.15}{473.15}~\si{\kelvin} at pressures up to \SI{30}{\mega\pascal} by Schilling et al.\ \cite{SchillingKleinrahmWagner:2008:1}.
Due to its good representation of liquid-phase experimental data, further works, \eg, of Sommer et al.\ \cite{SommerKleinrahmSpanWagner:2011:1}, Sampson et al.\ \cite{SampsonYangXuRichter:2019:1} (\cref{tbl:model_ethylene-glycol}, propylene glycol - fix Exp.) and Yang et al.\ \cite{YangSampsonFrotscherRichter:2020:1} (\cref{tbl:model_ethylene-glycol}, ethylene glycol - fix Exp.), also relate to this equation.
The so-called Schilling-type equation has the following functional form:
\begin{equation}
	\frac{\rho}{\rho_0}=\sum_{j=1}^{I_\textup{Pol}} n_j \, \sigma^{t_j} \pi^{p_j}
	\label{eq:Schilling}
\end{equation}
\begin{equation}
	\sigma=(T/T_0-1) \text{ and } \pi=(p/p_0+1)
\end{equation}
The parameters \(\sigma\) and \(\pi\) represent the reduced temperature and the reduced pressure, respectively, while \(T_0\), \(p_0\) and \(\rho_0\) are chosen by the measured values and \(I_\textup{{Pol}}\) defines the number of terms.
In Sampson et al.\ \cite{SampsonYangXuRichter:2019:1}, the parameters were set to \(T_0 =\)~\SI{150}{\kelvin}, \(p_0=\)~\SI{100}{\mega\pascal} and \(\rho_0=\)~\SI{1000}{\kilo\gram\per\meter\cubed}, and \(n_j\), \(t_j\) and \(p_j\) were fitted to the experimental data of propylene glycol setting the number of terms \(I_\textup{{Pol}}=\)~\num{8}.
With this model, the measured densities for propylene glycol can be reproduced within a maximum error of  \(\varepsilon_\textup{R}=0.015~\%\).
Due to the similarity between both substances, the same setup was used to fit the parameters \(n_j\) for ethylene glycol in Yang et al.\ \cite{YangSampsonFrotscherRichter:2020:1}, keeping the exponents \(t_j\) and \(p_j\) fixed (fixed exponent model, \cref{tbl:model_ethylene-glycol}).
The experimental data of ethylene glycol is represented within a maximum error of \(\varepsilon_\textup{R}=0.025~\%\).

Most current modeling approaches, especially for non-linear models, have a high development effort and the requirement of thorough background knowledge in common.
In the work of Yang et al.\ also a new approach utilizing Eureqa was employed to investigate the suitability of a machine learning tool for non-linear EOS modeling \cite{SchmidtLipson:2009:1,Eureqa:2015}.
The form of terms was chosen equivalent to the existing model, while the aim was to reduce the number of terms, to fit the exponents and to stay within an adequate uncertainty (Eureqa model, \cref{tbl:model_ethylene-glycol}). 
In addition to the less complex functional form, a maximum error of \(\varepsilon_\textup{R}=0.020~\%\) and a better extrapolation behavior are achieved.
\cref{tbl:compare_EOS-types} gives an overview of the different modeling approaches.

\begin{table*}[htp]
	\begin{center}
	\caption{Features of different EOS modeling approaches for liquid-phase densities (Tait equation \cite{GibsonLoeffler:1939:1}, Schilling-type equation \cite{SchillingKleinrahmWagner:2008:1}, Eureqa modeling \cite{SchmidtLipson:2009:1,Eureqa:2015})}
		\begin{tabular}{c|l}
			\hline
			EOS type & Feature         \\ \hline
			\multirow{3}{*}{Tait} 
				& well-established model, widely used for liquid densities \\
				& predefined model form \\
				& requires custom fitting algorithm \\
				\hline
			\multirow{3}{*}{Schilling} 
				& simple model form, related to fundamental EOS \\
				& user-specific number of terms \\ 
				& requires custom fitting algorithm \\
				\hline
			\multirow{3}{*}{Eureqa} 
				& intuitive software handling \\
				& symbolic regression based on genetic algorithms \\
				& comparing complexity and uncertainty \\
				\hline
		\end{tabular}
	\label{tbl:compare_EOS-types}
	\end{center}
\end{table*}

%% file: 03_background_OED.tex
Optimal experimental design is a technique to select experiments which are most informative about the unknown parameters of a given model.
We refer the interested reader to Pázman, Uciński and Atkinson et al.\ \cite{Pazman:1986:1,Ucinski:2005:1,AtkinsonDonevTobias:2007:1} for a comprehensive background.
Here, we use OED to reduce the experimental effort required to identify the parameters $n_j$, $j = 1, \ldots, I_\textup{{Pol}}$, in \eqref{eq:Schilling} with exponents $t_j$ and $p_j$ fixed.
In this case, the dependent variable $\rho/\rho_0$ is a linear function of the parameters, which simplifies the subsequent description.
To be specific, we deduce from \eqref{eq:Schilling} that each triple of measured data $(p_i, \rho_i, T_i)$, after conversion to reduced quantities $(\pi_i, \rho_i/\rho_0, \sigma_i)$, contributes a prediction of the form $\rho_i \approx \rho_0 \, j(\pi_i,\sigma_i) \, n$, where $n = (n_1, \ldots, n_8)^\transp \in \R^8$ denotes the parameter vector and 
\begin{equation}
	\label{eq:elementary_Jacobian}
	j(\pi_i,\sigma_i)
	=
	\begin{bmatrix}
		\sigma_i^{t_1} \pi_i^{p_1} & \cdots & \sigma_i^{t_8} \pi_i^{p_8}
	\end{bmatrix}
\end{equation}
is the so-called elementary Jacobian associated with the $i$-th measurement.
Following the theory of OED, the information content of a single measurement is expressed through the elementary Fisher information matrix (FIM) $I_i = j(\pi_i, \sigma_i)^\transp j(\pi_i, \sigma_i) \in \R^{8 \times 8}$.
Each elementary FIM is a symmetric, positive semi-definite rank-1 matrix.
Moreover, since we can assume individual measurements to be statistically independent, the FIM associated with a series of experiments is obtained as $I = \sum_i I_i$.

It is common to apply a scalar objective function, which converts the FIM associated with any collection of experiments into a single number and, thus, allows a comparison with any other collection of experiments.
We utilize here the A-criterion 
\begin{equation}
	\label{eq:OED_objective}
	\Psi_\textup{A}(I) 
	=
	\trace(I^{-1}) = \sum_{j=1}^8 \frac{1}{\lambda_j}, 
\end{equation}
where $\lambda_j$ is the $j$-th eigenvalue of $I$.
Clearly, at least eight individual measurements are required to render the FIM positive definite and the criterion $\Psi_\textup{A}(I)$ finite.
In this case, the value of $\Psi_\textup{A}(I)$ is proportional to the sum of the squared semi-axes of confidence ellipsoids in the 8-dimensional parameter space.
Consequently, we seek to minimize $\Psi_\textup{A}(I)$, possibly subject to constraints on the experimental budget, in order to maximize the information content of the collection of experiments selected and simultaneously minimize parameter variation in the face of measurement errors.

As we argued in the introduction, it is advantageous from a practical point of view to take measurements along an isotherm. 
In the following section, we will therefore optimize over experiments each of which comprises several measurements obtained by varying the pressure along an isotherm.

%% file: 04_numerical_results.tex
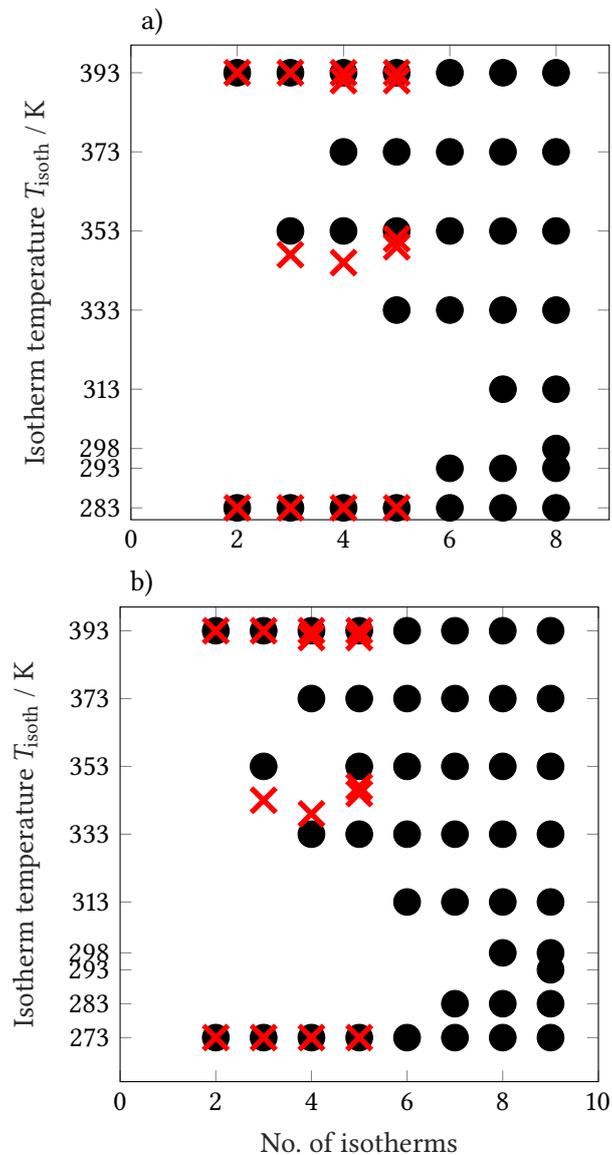
\begin{figure}[htp]
	\centering
	\begin{subfigure}{\linewidth}
		\centering
		\input{matlab/tikz_best_measIsot.tex}
	\end{subfigure}
	\begin{subfigure}{\linewidth}
		\centering
		\input{matlab/tikz_best_measIsot_propgly.tex}
	\end{subfigure}
	\caption{Selection of the best isotherms \(\bullet\) from the measured isotherms and \textcolor{red}{x} from a free choice a) for ethylene glycol between \(T = \)~(\numrange{283}{393})~\si{\kelvin} and b) for propylene glycol between \(T = \)~(\numrange{273}{393})~\si{\kelvin}, within a raster of \SI{2}{\kelvin} using the A-criterion \eqref{eq:OED_objective} to minimize the parameter uncertainty}
	\label{fig:bestIsot}
\end{figure}

The approach described in \cref{section:oed} is used to optimize the measurement series by maximizing the information contained in a set of eight isothermal measurements at approximate temperatures $T = $~(\numlist{283; 293; 298; 313; 333; 353; 373; 393})~\si{\kelvin} drawn from the experiments conducted by Yang et al.\ \cite{YangSampsonFrotscherRichter:2020:1}.
The measurement plan for each isotherm consists of nine pressure values, approximately $p = $~(\numlist{5; 10; 15; 20; 30; 50; 70; 90; 100})~\si{\mega\pascal}.
\cref{fig:bestIsot}~a) and \cref{tbl:bestIsot_objective} show the most informative selection of isotherms (black dots) to fit the parameters \(n_1, \ldots, n_8\) using a varying number of isotherms.
As expected, the value of the objective \eqref{eq:OED_objective} decreases as we allow more isotherms to be included.
However, when transitioning from five to six isotherms, the further decrease in the objective is small compared to the previous steps.
For propylene glycol, measured at approximate temperatures $T = $~(\numlist{273; 283; 293; 298; 313; 333; 353; 373; 393})~\si{\kelvin}, \cref{fig:bestIsot}~b) and \cref{tbl:bestIsot_objective_PG} show a similar selection of best isotherms and for the decay of the objective function values.
In this case, the measurement plan for each isotherm consists of eight pressure values, approximately $p = $~(\numlist{5; 10; 15; 20; 30.5; 50.5; 71; 91})~\si{\mega\pascal}.

\begin{table*}[htb]
	\begin{center}
		\caption{Selection of the best isotherms for different numbers of isotherms and the objective value for ethylene glycol}
		\sisetup{round-mode=places,round-precision=2, table-format = 1.1e1}
		\setlength\extrarowheight{2pt}
		\resizebox{\textwidth}{!}{%
			\begin{tabular}{c|l|r}
				\hline
				No.\ of & measured isotherms & objective \\
				isotherms & best choice \si{\kelvin} & value \\ 
				\hline
				2 & \numlist{393.09; 283.32} & \num{23782.8693013232} \\
				3 & \numlist{393.09; 283.32; 353.13} & \num{120.317879170357} \\
				4 & \numlist{393.09; 283.32; 353.13; 373.22} & \num{94.4000855869682} \\
				5 & \numlist{393.09; 283.32; 353.13; 373.22; 333.09} & \num{78.8200645162156}	\\
				6 & \numlist{393.09; 283.32; 353.13; 373.22; 333.09; 293.14} & \num{73.0657921007370} \\
				7 & \numlist{393.09; 283.32; 353.13; 373.22; 333.09; 293.14; 313.13} & \num{68.3607110877593}	\\
				8 & \numlist{393.09; 283.32; 353.13; 373.22; 333.09; 293.14; 313.13; 298.18} & \num{65.7744767992581} \\ 
				\hline 
				\hline
				No.\ of & free isotherms & objective \\ 
				isotherms & best choice \si{\kelvin} & value \\ 
				\hline
				2 & \numlist{283.00; 393.00} & \num{23113.1296526466} \\
				3 & \numlist{283.00; 347.00; 393.00} & \num{119.827863581756} \\
				4 & \numlist{283.00; 345.00; 391.00; 393.00} & \num{85.9831108316136}\\
				5 & \numlist{283.00; 349.00; 351.00; 391.00; 393.00} & \num{69.2247} \\ 
				\hline
			\end{tabular}
		}
		\label{tbl:bestIsot_objective}
	\end{center}
\end{table*}

\begin{table*}[htb]
	\begin{center}
		\caption{Selection of the best isotherms for different numbers of isotherms and the objective value for propylene glycol}
		\sisetup{round-mode=places,round-precision=2, table-format = 1.1e1}
		\setlength\extrarowheight{2pt}
		\resizebox{\textwidth}{!}{%
			\begin{tabular}{c|l|r}
				\hline
				No.\ of & measured isotherms & objective \\
				isotherms & best choice \si{\kelvin} & value \\ 
				\hline
				2 & \numlist{272.73; 392.95} & \num{37026.6003363935} \\
				3 & \numlist{272.73; 392.95; 352.99} & \num{145.843096759086} \\
				4 & \numlist{272.73; 392.95; 373.34; 333.05} & \num{107.347985002866} \\
				5 & \numlist{272.73; 392.95; 352.99; 373.34; 333.05} & \num{90.8605150637882}	\\
				6 & \numlist{272.73; 392.95; 352.99; 373.34; 333.05; 313.12} & \num{83.2598627774449} \\
				7 & \numlist{272.73; 392.95; 352.99; 373.34; 333.05; 313.12; 283.18} & \num{78.5644530295377}	\\
				8 & \numlist{272.73; 392.95; 352.99; 373.34; 333.05; 313.12; 283.18; 298.12} & \num{75.5465916354905} \\ 
				9 & \numlist{272.73; 392.95; 352.99; 373.34; 333.05; 313.12; 283.18; 298.12; 293.18} & \num{73.4700479773580} \\
				\hline 
				\hline
				No.\ of & free isotherms & objective \\ 
				isotherms & best choice \si{\kelvin} & value \\ 
				\hline
				2 & \numlist{273.00; 393.00} & \num{35604.6999516390} \\
				3 & \numlist{273.00; 343.00; 393.00} & \num{144.110407340345} \\
				4 & \numlist{273.00; 339.00; 391.00; 393.00} & \num{98.7752993364498}\\
				5 & \numlist{273.00; 345.00; 347.00; 391.00; 393.00} & \num{80.8285849746501} \\ 
				\hline
			\end{tabular}
		}
		\label{tbl:bestIsot_objective_PG}
	\end{center}
\end{table*}

We aim to compare the accuracies of the fixed exponent model \eqref{eq:Schilling} for ethylene glycol, once with parameters $n_1, \ldots, n_8$ fitted from the complete set of experiments and once using only the five best isotherms (\cref{tbl:model_ethylene-glycol} fixE-5).
To this end, we show in \cref{fig:relError}~a) and c) the relative deviations of calculated densities from experimental values as a function of pressure along all eight isotherms. It can be seen that using only five out of eight isotherms for the fitting process yields a greatest deviation only larger by approximately $\SI{0.003}{\%}$.
The same approach is used to fit a new model for propylene glycol to the best five isotherms, see the lower part of \cref{tbl:model_ethylene-glycol} (fixE-5).
With an absolute difference in the greatest deviation of $\SI{0.0015}{\%}$ for propylene glycol between the two fixed exponent models (\cref{fig:relError_PG}~a) and b)), the result for OED is even more promising than for ethylene glycol.

Next, we investigate the information gain obtained by allowing the temperatures of the isothermal experiments to be chosen freely within the interval from $T = \SI{283}{\kelvin}$ to $T = \SI{393}{\kelvin}$ for ethylene glycol and $T = \SI{273}{\kelvin}$ to $T = \SI{393}{\kelvin}$ for propylene glycol.
To this end, we introduce an equidistant grid of possible temperature values with a spacing of $\SI{2}{\kelvin}$, and pressure values for each group of isothermal experiments are the same as measured.
The optimal selections of up to five isotherms can be found in \cref{fig:bestIsot}~a) and the bottom half of \cref{tbl:bestIsot_objective} for ethylene glycol as well as in \cref{fig:bestIsot}~b) and the bottom half of \cref{tbl:bestIsot_objective_PG} for propylene glycol.

By comparing both selection approaches, the similarity of temperatures for the best two and three isotherms as well as the difference between those for the four and five best isotherms are noticeable.
This leads us to expect an improvement of the model with a fit to the free best five isotherms.
However, due to the missing measured values, no comparable model can be developed for this case.

Instead, we investigate the importance of the functional form by fitting a new model to the best five measured isotherms, where also the exponents are optimized (\cref{fig:relError}~d) and \cref{tbl:model_ethylene-glycol} freeE-5).
Here, even with five isotherms, the maximum deviations can be reduced significantly.
However, for propylene glycol, the free exponent model shows large deviations for the isotherms not selected (greatest deviation $\SI{0.0431}{\%}$) compared to the deviation of the selected isotherms in the free exponent model (greatest deviation $\SI{0.0046}{\%}$) (\cref{fig:relError_PG}~c)) and compared to the model published by Sampson et al \cite{SampsonYangXuRichter:2019:1} (\cref{tbl:model_ethylene-glycol}, propylene glycol - fix Exp.).
This behavior in conjunction with the conclusion of Yang et al.\ \cite{YangSampsonFrotscherRichter:2020:1} concerning the Eureqa model (\cref{fig:relError}~b)), demonstrates the importance of a comprehensive model optimization.
To exploit the potential of free exponent modeling, we plan to employ the sequential OED approach, described in \cref{section:introduction}, for calculating the necessary isotherms based on nonlinear models in future work.

\begin{figure}[htp]
	\centering
	\begin{subfigure}{\linewidth}
		\centering
		\input{matlab/tikz_relError_fixedExp.tex}
	\end{subfigure}
	\\
	\begin{subfigure}{\linewidth}
		\centering
		\input{matlab/tikz_relError_Eureqa.tex}
	\end{subfigure}
	\\
	\rput(0.1,5){~~ \rotatebox{90}{~~~~~~~~~~~~~~~~~~~~~~~~~~~~~~~~~~~~~~$\text{100}\cdot\text{(}\rho{}_{\text{calc}}-\rho{}_{\text{exp}}\text{)/}\rho{}_{\text{exp}}$ / \si{\%}}}%
	\begin{subfigure}{\linewidth}
		\centering
		\input{matlab/tikz_relError_fixedExp_5measIsot.tex}
	\end{subfigure}
	\\
	\begin{subfigure}{\linewidth}
		\centering
		\input{matlab/tikz_relError_freeExp_5measIsot.tex}
	\end{subfigure}
	\sisetup{round-mode=places,round-precision=0}
	\caption{Relative deviations of \textbf{ethylene glycol} densities calculated with different models from experimental values, a) fixed exponent model fitted with all eight measured isotherms, b) Eureqa model fitted with all eight measured isotherms, c) fixed exponent model fitted with the best five measured isotherms (bold marker), d) free exponent model fitted with the best five measured isotherms (bold marker). \textcolor{black}{\rotatebox[origin=c]{180}{\(\triangle\)}}, \(T\approx\)\,\SI{283.32}{\kelvin}; \textcolor{mycolor1}{\(+\)}, \(T\approx\)\,\SI{293.142}{\kelvin}; \textcolor{mycolor1}{\rotatebox[origin=c]{270}{\(\triangle\)}}, \(T\approx\)\,\SI{293.140}{\kelvin}; \textcolor{mycolor2}{o}, \(T\approx\)\,\SI{298.18}{\kelvin}; \textcolor{mycolor3}{\(*\)}, \(T\approx\)\,\SI{313.13}{\kelvin}; \textcolor{mycolor4}{x}, \(T\approx\)\,\SI{333.09}{\kelvin}; \textcolor{mycolor5}{\(\boxempty\)}, \(T\approx\)\,\SI{353.13}{\kelvin}; \textcolor{mycolor6}{\(\diamond\)}, \(T\approx\)\,\SI{373.22}{\kelvin}; \textcolor{mycolor7}{\(\triangle\)}, \(T\approx\)\,\SI{393.09}{\kelvin}}
	\label{fig:relError}
\end{figure}
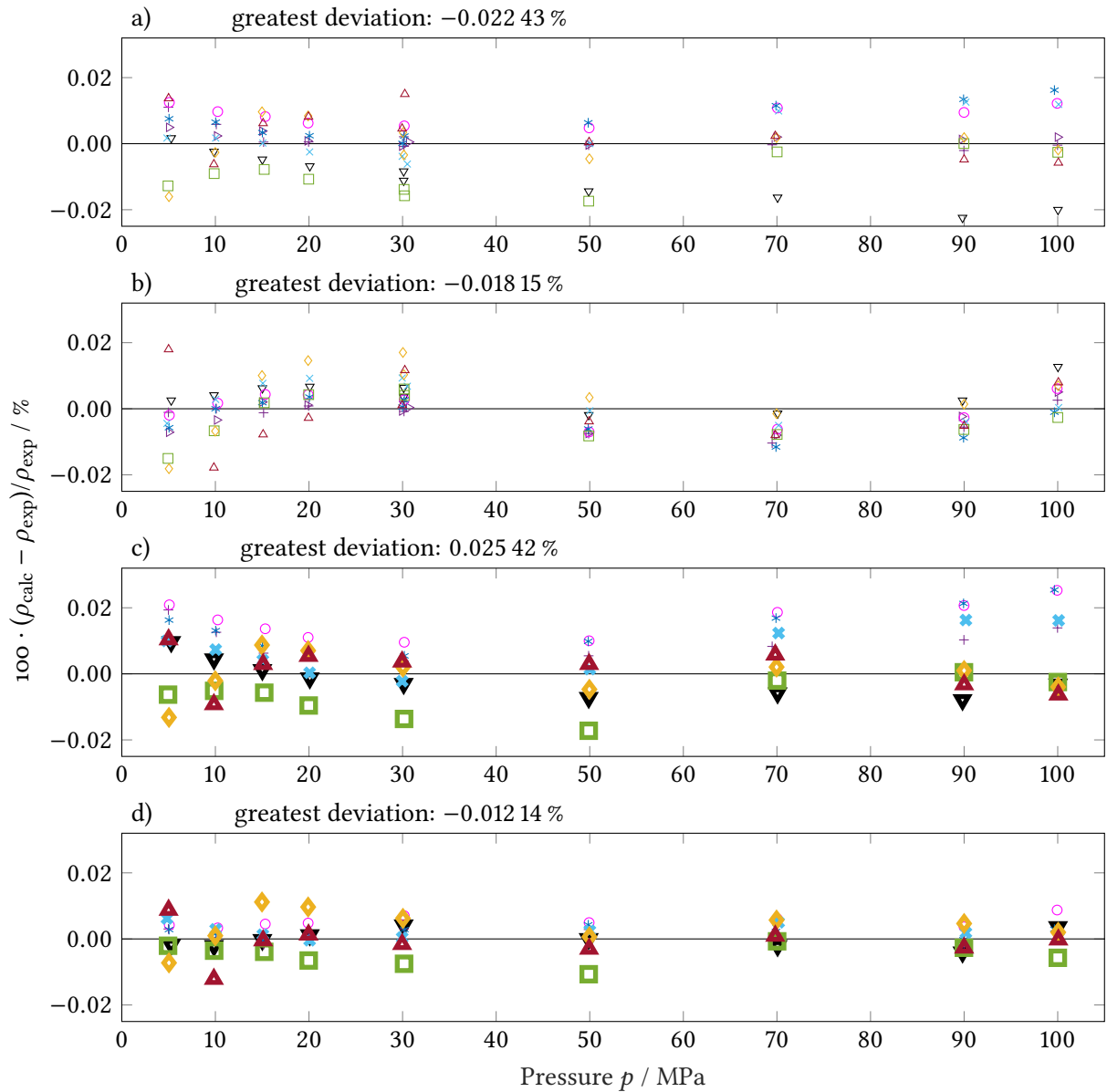
\begin{figure}[htp]
	\centering
	\begin{subfigure}{\linewidth}
		\centering
		\input{matlab/tikz_relError_fixedExp_9measIsot_PG.tex}
	\end{subfigure}
	\\
	\rput(0.1,5){~~ \rotatebox{90}{$\text{100}\cdot\text{(}\rho{}_{\text{calc}}-\rho{}_{\text{exp}}\text{)/}\rho{}_{\text{exp}}$ / \si{\%}}}%
	\begin{subfigure}{\linewidth}
		\centering
		\input{matlab/tikz_relError_fixedExp_5measIsot_PG.tex}
	\end{subfigure}
	\\
	\begin{subfigure}{\linewidth}
		\centering
		\input{matlab/tikz_relError_freeExp_5measIsot_PG.tex}
	\end{subfigure}
	\sisetup{round-mode=places,round-precision=0}
	\caption{Relative deviations of \textbf{propylene glycol} densities calculated with different models from experimental values, a) fixed exponent model fitted with all nine measured isotherms, b) fixed exponent model fitted with the best five measured isotherms (bold marker), c) free exponent model fitted with the best five measured isotherms (bold marker). \textcolor{black}{\rotatebox[origin=c]{180}{\(\triangle\)}}, \(T\approx\)\,\SI{272.73}{\kelvin}; \textcolor{mycolor1}{\(+\)}, \(T\approx\)\,\SI{283.18}{\kelvin}; \textcolor{mycolor2}{o}, \(T\approx\)\,\SI{293.18}{\kelvin}; \textcolor{mycolor3}{\(*\)}, \(T\approx\)\,\SI{298.12}{\kelvin}; \textcolor{mycolor4}{x}, \(T\approx\)\,\SI{313.12}{\kelvin}; \textcolor{mycolor5}{\(\boxempty\)}, \(T\approx\)\,\SI{333.05}{\kelvin}; \textcolor{mycolor6}{\(\diamond\)}, \(T\approx\)\,\SI{352.99}{\kelvin}; \textcolor{mycolor8}{\rotatebox[origin=c]{90}{\(\triangle\)}}, \(T\approx\)\,\SI{373.34}{\kelvin}; \textcolor{mycolor7}{\(\triangle\)}, \(T\approx\)\,\SI{392.95}{\kelvin}}
	\label{fig:relError_PG}
\end{figure}
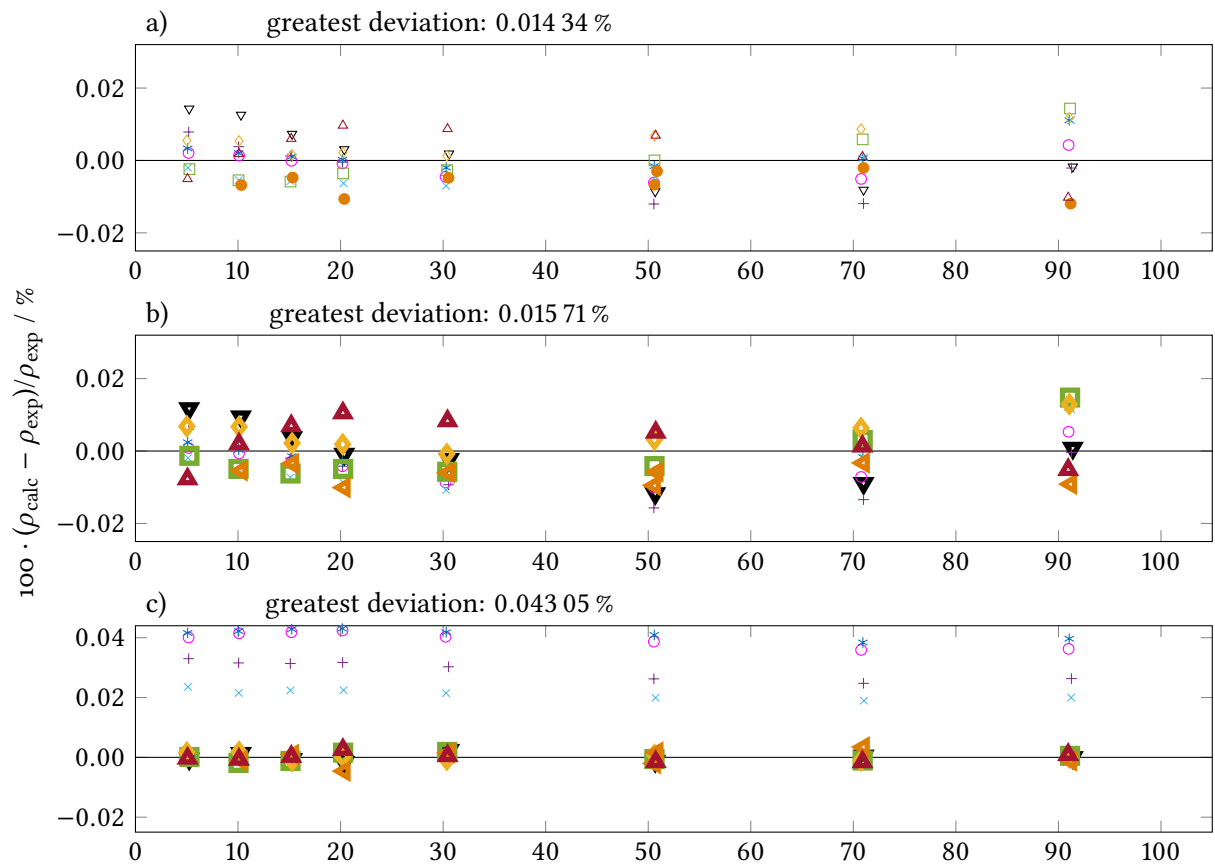

\begin{table*}[htp]
	\caption{Coefficients of \eqref{eq:Schilling} for the fixed exponent model (fix. Exp.), the Eureqa model  \cite{YangSampsonFrotscherRichter:2020:1}, fixed exponent model fitted with the best five measured isotherms (fixE-5) and free exponent model fitted with the best five measured isotherms (densities calculated by the Eureqa model) (freeE-5) for ethylene glycol and propylene glycol}
	\sisetup{round-mode=places,round-precision=3,table-format = 11.1e1,round-integer-to-decimal,scientific-notation=true}
	\setlength\extrarowheight{2pt}
	\resizebox{\textwidth}{!}{%
		\begin{tabular}{cc|cccccccc}
			\hline
			\multicolumn{10}{c}{\textbf{ethylene glycol}} \\ \hline
			& \(i\)    & 1         & 2         & 3         & 4         & 5         & 6         & 7         & 8         \\ \hline
			\parbox[t]{2mm}{\multirow{3}{*}{\rotatebox[origin=c]{90}{fix Exp.}}} & \(n_i\) & \num{1.21179} & \num{-0.115946} & \num{-0.00229491} & \num{0.0167979} & \num{-0.00642017} & \num{0.00319735} & \num{-0.0000562648} & \num{-0.000104477} \\
			& \(t_i\) & \(0.0\)  & \(1.0\)  & \(4.0\)  & \(2.5\)  & \(3.0\) & \(-0.5\)  & \(3.5\) & \(5.5\) \\
			& \(p_i\) & \(0.0\) & \(-0.5\) & \(-2.5\) & \(-4.0\) & \(-6.0\) & \(2.0\) & \(2.5\) & \(0.5\)  \\ \hline
			\parbox[t]{2mm}{\multirow{3}{*}{\rotatebox[origin=c]{90}{Eureqa}}} & \(n_i\) & \num{1.07435} & \num{0.0263317} & \num{0.0429109} & \num{-0.00389781} & \num{0.00253998} & \num{-0.0100758} & \num{-0.0254439} & -         \\
			& \(t_i\) & \(0.0\)  & \(0.0\)    & \(-1.0\)         & \(-1.0\)         & \(2.0\)         & \(2.0\)         & \(2.0\)         & -         \\
			& \(p_i\) & \(0.0\)  & \(1.0\)   & \(0.0\)        & \(-1.0\)        & \(1.0\)       & \(-1.0\)         & \(0.0\)         & -         \\ \hline
			\parbox[t]{2mm}{\multirow{3}{*}{\rotatebox[origin=c]{90}{fixE-5}}} & \(n_i\) & \num{1.21163360445923}  & \num{-0.115905514331211} & \num{-0.00253163502333771} & \num{0.0172128144265885} & \num{-0.00645755779627614} & \num{0.00329061192745139} & \num{-8.07047465004974e-05} & \num{-6.64736081540256e-05} \\
			& \(t_i\) & \(0.0\)  & \(1.0\)  & \(4.0\)  & \(2.5\)  & \(3.0\) & \(-0.5\)  & \(3.5\) & \(5.5\) \\
			& \(p_i\) & \(0.0\) & \(-0.5\) & \(-2.5\) & \(-4.0\) & \(-6.0\) & \(2.0\) & \(2.5\) & \(0.5\)  \\ \hline
			\parbox[t]{2mm}{\multirow{3}{*}{\rotatebox[origin=c]{90}{freeE-5 }}} & \(n_i\) & \num{1.12982703408683}  & \num{-0.0307811706964866} & \num{-0.000317796487568905} & \num{0.0100711779206958} & \num{-0.00372120778782193} & \num{0.00200502111639758} & \num{-4.64941130891909e-05} & \num{-0.000150990620106622} \\
			& \(t_i\) & \num{-0.0474071486759451}  & \num{2.16387158399161}  & \num{3.90224775360683}  & \num{3.51853884268139}  & \num{4.45963786135804} & \num{-0.359833567281470}  & \num{3.49246747472542} & \num{4.82949327242230} \\
			& \(p_i\) & \num{0.0217278770574652} & \num{-1.04784849658914} & \num{-2.49167127531736} & \num{-4.44248295580573} & \num{-6.17332350298351} & \num{2.41222619454673} & \num{2.47595618623397} & \num{0.685392927836445}  \\ \hline \hline
			\multicolumn{10}{c}{\textbf{propylene glycol}} \\ \hline
			& \(i\)    & 1         & 2         & 3         & 4         & 5         & 6         & 7         & 8         \\ \hline
			\parbox[t]{2mm}{\multirow{3}{*}{\rotatebox[origin=c]{90}{fix Exp.}}} & \(n_i\) & \num{11.4042000000000} & \num{-0.117385000000000} & \num{-0.00237453000000000} & \num{0.0103010000000000} & \num{-0.00307390000000000} & \num{0.00383355000000000} & \num{-2.59107000000000e-05} & \num{-0.000183493000000000} \\
			& \(t_i\) & \(0.0\)  & \(1.0\)  & \(4.0\)  & \(2.5\)  & \(3.0\) & \(-0.5\)  & \(3.5\) & \(5.5\) \\
			& \(p_i\) & \(0.0\) & \(-0.5\) & \(-2.5\) & \(-4.0\) & \(-6.0\) & \(2.0\) & \(2.5\) & \(0.5\)  \\ \hline
			\parbox[t]{2mm}{\multirow{3}{*}{\rotatebox[origin=c]{90}{fixE-5}}} & \(n_i\) & \num{1.14026018088541}  & \num{-0.117541083757401} & \num{-0.00245762014885811} & \num{0.0110445690096595} & \num{-0.00348027882689822} & \num{0.00389421244388974} & \num{-1.88557765207714e-05} & \num{-0.000183276381114324} \\
			& \(t_i\) & \(0.0\)  & \(1.0\)  & \(4.0\)  & \(2.5\)  & \(3.0\) & \(-0.5\)  & \(3.5\) & \(5.5\) \\
			& \(p_i\) & \(0.0\) & \(-0.5\) & \(-2.5\) & \(-4.0\) & \(-6.0\) & \(2.0\) & \(2.5\) & \(0.5\)  \\ \hline
			\parbox[t]{2mm}{\multirow{3}{*}{\rotatebox[origin=c]{90}{freeE-5 }}} & \(n_i\) & \num{5.79584320383811}  & \num{-4.77469983690109} & \num{-0.00143264642403525} & \num{0.00410372816596997} & \num{-0.00107279822860758} & \num{0.00448510890563661} & \num{0.00513977401208411} & \num{-0.000478994572062626} \\
			& \(t_i\) & \num{0.285926521936452}  & \num{0.373933291576127}  & \num{3.81315450449070}  & \num{2.25730639679115}  & \num{2.78297227903763} & \num{-0.657639394856825}  & \num{3.84032621482629} & \num{6.98387533348200} \\
			& \(p_i\) & \num{-0.0294108898980225} & \num{-0.0455971798229361} & \num{-2.68704589670844} & \num{-4.14016473675111} & \num{-6.08201830177042} & \num{1.64560575522704} & \num{0.353357536187274} & \num{0.258317219703911}  \\ \hline	
		\end{tabular}
	}
	\label{tbl:model_ethylene-glycol}
\end{table*}

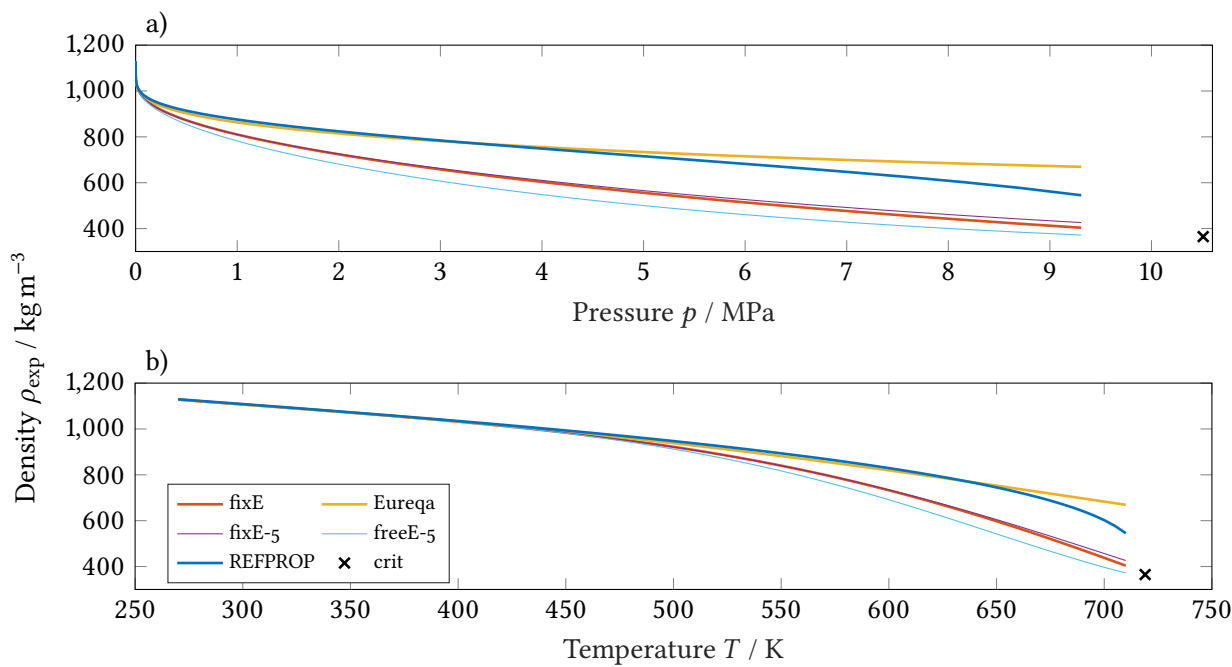
\begin{figure}[ht]
	\centering
	\begin{subfigure}{\linewidth}
		\centering
		\input{matlab/tikz_extrapol2VLsat_pressure.tex}
	\end{subfigure}
	\\
	\rput(0,0){~~ \rotatebox{90}{~~~~~~~~~~~~~~~~~~~~~~~~~~~~~~~~~~~~~ Density $\rho_{\text{exp}}$ / \si{\kilo\gram\per\meter\cubed}}}
	\begin{subfigure}{\linewidth}
		\centering
		\input{matlab/tikz_extrapol2VLsat_temperature.tex}
	\end{subfigure}
	\caption{Liquid densities at the vapor-liquid saturation line for a) the saturation pressure and b) the saturation temperature obtained from the fundamental equation embedded in Refprop, calculated with the fixed exponent model, the Eureqa model, the fixed exponents model, using the five best measured isotherms, the free exponent model fitted with the best five measured isotherms and the fundamental equation and the critical point embedded in REFPROP \cite{Refprop:2018}}
	\label{fig:extrSat}
\end{figure}

For a more comprehensive model examination between the modeling approach by Eureqa and by the typical Schilling-type fit, the extrapolation behavior was investigated taking the example of ethylene glycol.
Therefore, the models were compared to an unpublished fundamental equation of state in form of the Helmholtz energy as implemented in REFPROP \cite{Refprop:2018}.
This model is valid within the following limits:
\begin{itemize}
	\item 
	temperature: \(T = \)~\numrange{260.6}{750.0}~\si{\kelvin},
	\item 
	pressure: up to \(p = \)~\SI{100}{\mega \pascal},
	\item 
	density: up to \(\rho = \)~\SI{1136.5}{\kilo\gram\per\meter\cubed},
	\item 
	vapor-liquid saturation line.
\end{itemize}
The vapor-liquid saturation line defines the limit for extrapolation up to the maximum temperature and the minimum pressure.
To calculate the liquid densities with the different models at the saturation line, the corresponding temperatures and pressures from REFPROP (\cref{fig:extrSat}) are used.
As already shown in the paper from Yang et al.\ \cite{YangSampsonFrotscherRichter:2020:1}, the Eureqa model (\cref{fig:extrSat}) shows the best extrapolation behavior within \SI{\pm 2}{\%} up to \(p=\)~\SI{4.5}{\mega \pascal} and \(T=\)~\SI{655}{\kelvin} with \(\rho \approx \)~\SI{740}{\kilo\gram\per\meter\cubed}.
The graphs for the fixed exponent model and the fixed exponent model with the best five measured isotherms in \cref{fig:extrSat} are similar, while the free exponent model fitted to the best five measured isotherms shows the largest deviations from the densities calculated with the fundamental equation in REFPROP.
It is important to note that none of the models reproduces the curvature of densities calculated from the fundamental equation in REFPROP, which means that they are all strictly limited to the calculation of liquid-phase densities.

Many applications work at pressures below \SI{5}{\mega\pascal}.
For this reason, we have extrapolated the measured isotherms to ambient pressure. 
All models reproduce the values within relative deviations of \numrange{-0.28}{-0.1}~\si{\%}.
Each model shows the largest deviation at the \SI{393}{\kelvin} isotherm, and again, the Eureqa model yields the best result with a maximum deviation of \SI{-0.20}{\%}.
We aim to consider boundary conditions, such as the extrapolation behavior, in further investigations.

%% file: matlab/tikz_best_measIsot.tex
%
%
\begin{tikzpicture}
	
	\begin{axis}[%
		width=\numapdefigurewidth,
		height=\numapdefigurewidth,
		xmin=0,
		xmax=9,
		xlabel style={font=\color{white!15!black}},
		ymin=280,
		ymax=400,
		ytick={283, 293, 298, 313, 333, 353, 373, 393},
		ylabel style={font=\color{white!15!black}},
		ylabel={Isotherm temperature $T_{\textup{isoth}}$ / K},
		axis background/.style={fill=white},
		every axis title/.style={right,at={(-0.0,1.05)}},
		title={a)},
		]
		\addplot [color=black, only marks, mark size=5.0pt, mark=*, mark options={solid, fill=black, black}, forget plot]
		table[row sep=crcr]{%
			2	283\\
			2	393\\
			3	283\\
			3	353\\
			3	393\\
			4	283\\
			4	353\\
			4	373\\
			4	393\\
			5	283\\
			5	333\\
			5	353\\
			5	373\\
			5	393\\
			6	283\\
			6	293\\
			6	333\\
			6	353\\
			6	373\\
			6	393\\
			7	283\\
			7	293\\
			7	313\\
			7	333\\
			7	353\\
			7	373\\
			7	393\\
			8	283\\
			8	293\\
			8	298\\
			8	313\\
			8	333\\
			8	353\\
			8	373\\
			8	393\\
		};
		\addplot [color=red, line width=2.0pt, only marks, mark size=6.5pt, mark=x, mark options={solid, fill=red, red}, forget plot]
		table[row sep=crcr]{%
			2	283\\
			2	393\\
			3	283\\
			3	347\\
			3	393\\
			4	283\\
			4	345\\
			4	391\\
			4	393\\
			5	283\\
			5	349\\
			5	351\\
			5	391\\
			5	393\\
		};
	\end{axis}
\end{tikzpicture}%

%% file: matlab/tikz_best_measIsot_propgly.tex
%
%
\begin{tikzpicture}

\begin{axis}[%
	width=\numapdefigurewidth,
	height=\numapdefigurewidth,
	xmin=0,
	xmax=10,
	xlabel style={font=\color{white!15!black}},
	xlabel={No.\ of isotherms},
	ymin=260,
	ymax=400,
	ytick={273, 283, 293, 298, 313, 333, 353, 373, 393},
	ylabel style={font=\color{white!15!black}},
	ylabel={Isotherm temperature $T_{\textup{isoth}}$ / K},
	axis background/.style={fill=white},
	every axis title/.style={right,at={(-0.0,1.05)}},
	title={b)},
	]
\addplot [color=black, only marks, mark size=5.0pt, mark=*, mark options={solid, fill=black, black}, forget plot]
table[row sep=crcr]{%
	2	273\\
	2	393\\
	3	273\\
	3	353\\
	3	393\\
	4	273\\
	4	333\\
	4	373\\
	4	393\\
	5	273\\
	5	333\\
	5	353\\
	5	373\\
	5	393\\
	6	273\\
	6	313\\
	6	333\\
	6	353\\
	6	373\\
	6	393\\
	7	273\\
	7	283\\
	7	313\\
	7	333\\
	7	353\\
	7	373\\
	7	393\\
	8	273\\
	8	283\\
	8	298\\
	8	313\\
	8	333\\
	8	353\\
	8	373\\
	8	393\\
	9	273\\
	9	283\\
	9	293\\
	9	298\\
	9	313\\
	9	333\\
	9	353\\
	9	373\\
	9	393\\
	};
\addplot [color=red, line width=2.0pt, only marks, mark size=6.5pt, mark=x, mark options={solid, fill=red, red}, forget plot]
table[row sep=crcr]{%
	2	273\\
	2	393\\
	3	273\\
	3	343\\
	3	393\\
	4	273\\
	4	339\\
	4	391\\
	4	393\\
	5	273\\
	5	345\\
	5	347\\
	5	391\\
	5	393\\
};
\end{axis}
\end{tikzpicture}%

%% file: matlab/tikz_relError_fixedExp.tex
%
%
\definecolor{mycolor1}{rgb}{0.49400,0.18400,0.55600}%
\definecolor{mycolor2}{rgb}{1.00000,0.00000,1.00000}%
\definecolor{mycolor3}{rgb}{0.00000,0.44700,0.74100}%
\definecolor{mycolor4}{rgb}{0.30100,0.74500,0.93300}%
\definecolor{mycolor5}{rgb}{0.46600,0.67400,0.18800}%
\definecolor{mycolor6}{rgb}{0.92900,0.69400,0.12500}%
\definecolor{mycolor7}{rgb}{0.63500,0.07800,0.18400}%
\begin{tikzpicture}[]
\node[] at (4,3) {greatest deviation: \SI[round-mode=figures, round-precision=4]{-0.0224307347856036}{\%}};
\begin{axis}[%
width=\textwidth,
height=1.7in,
xmin=0,
xmax=105,
ymin=-0.025,
ymax=0.032,
ylabel style={font=\color{white!15!black}, text opacity=0},
ylabel={$\text{100}\cdot\text{(}\rho{}_{\text{exp}}\text{)/}\rho{}_{\text{exp}}$ / \si{\%}},
scaled ticks=false,
y tick label style={/pgf/number format/.cd,fixed,precision=2,zerofill},
axis background/.style={fill=white},
every axis title/.style={right,at={(-0.0,1.1)}},
title={a)},
]
\addplot [color=black, forget plot]
  table[row sep=crcr]{%
0	0\\
110	0\\
};
\addplot [color=black, only marks, mark=triangle, mark options={solid, rotate=180, black}, forget plot]
  table[row sep=crcr]{%
5.28	0.00166777887856839\\
9.85	-0.00241317748978736\\
15.02	-0.00477802877743516\\
20.09	-0.00682124505210324\\
30.12	-0.00837280512036559\\
49.89	-0.0144132731645188\\
70.07	-0.0162990069315673\\
89.83	-0.0224307347856036\\
100.03	-0.0200331463202587\\
30.12	-0.0112020654200789\\
};
\addplot [color=mycolor1, only marks, mark=+, mark options={solid, mycolor1}, forget plot]
  table[row sep=crcr]{%
4.98	0.011035867875062\\
10.1	0.00583936452967832\\
15.15	0.00049291363719715\\
19.96	0.000533895218219775\\
30.15	-0.000810967006747206\\
49.88	-0.000423806832462217\\
69.49	-0.000332316845067113\\
89.98	-0.00207431147519734\\
99.98	-0.000461916263733418\\
30.13	0.0020959481891504\\
};
\addplot [color=mycolor1, only marks, mark=triangle, mark options={solid, rotate=270, mycolor1}, forget plot]
table[row sep=crcr]{%
5.04	0.00493558602500362\\
10.16	0.0023440483090138\\
15	0.00381346797926277\\
19.84	0.000850599815726765\\
29.94	-0.000737262114270697\\
49.86	-0.000384383867085143\\
69.97	0.00201516432302168\\
89.75	0.0012765258014704\\
100	0.00198103615868559\\
30.68	0.000495758072319364\\
};
\addplot [color=mycolor2, only marks, mark=o, mark options={solid, mycolor2}, forget plot]
  table[row sep=crcr]{%
5.08	0.0124993575442409\\
10.25	0.00970602636858328\\
15.33	0.00819766043869058\\
19.91	0.00623773571662552\\
30.18	0.00538943161904067\\
49.93	0.00477301198579703\\
70.05	0.0108009059534363\\
89.99	0.00946391108395474\\
99.94	0.0122006526847638\\
};
\addplot [color=mycolor3, only marks, mark=asterisk, mark options={solid, mycolor3}, forget plot]
  table[row sep=crcr]{%
5.05	0.00754802847374698\\
10.03	0.00650917302936167\\
15.02	0.0033780736648664\\
20.05	0.00240845654968276\\
30.19	0.00230659184165988\\
49.84	0.00633083251404452\\
69.9	0.01149918081135\\
89.94	0.0134276009095368\\
99.65	0.0162134927745192\\
30	8.81117457547671e-05\\
};
\addplot [color=mycolor4, only marks, mark=x, mark options={solid, mycolor4}, forget plot]
  table[row sep=crcr]{%
4.83	0.00171205084030386\\
10.03	0.00171003868461607\\
15.07	0.000198843816199037\\
20.07	-0.00250082266602247\\
29.96	-0.00382857655357317\\
50.01	4.35225674698421e-07\\
70.18	0.00990901313684894\\
90.19	0.0125777051261512\\
100.1	0.0118914726129611\\
30.51	-0.00617091548299375\\
};
\addplot [color=mycolor5, only marks, mark=square, mark options={solid, mycolor5}, forget plot]
  table[row sep=crcr]{%
4.94	-0.0127723532340299\\
9.87	-0.00905686075167441\\
15.22	-0.00780283407181416\\
19.97	-0.0107543946004715\\
30.16	-0.0138361541946962\\
49.89	-0.0173982881430068\\
70.02	-0.00248916637221934\\
89.98	5.01291920522968e-05\\
100.01	-0.00260847061885505\\
30.21	-0.0157698131067363\\
};
\addplot [color=mycolor6, only marks, mark=diamond, mark options={solid, mycolor6}, forget plot]
  table[row sep=crcr]{%
5.04	-0.0160504610627454\\
9.99	-0.00264838209534168\\
14.98	0.00963878504611566\\
19.91	0.00841408666129362\\
30.04	0.00334244953763969\\
49.96	-0.00457810093054436\\
69.96	0.00183904246641199\\
90	0.00184266609414253\\
100.05	-0.00190670711316405\\
30.14	-0.00341575656096407\\
};
\addplot [color=mycolor7, only marks, mark=triangle, mark options={solid, mycolor7}, forget plot]
  table[row sep=crcr]{%
5	0.0137263413640288\\
9.84	-0.00625772681965626\\
15.1	0.00617906431531724\\
19.95	0.00813937681320096\\
29.95	0.00456597869431767\\
49.93	0.000473426568496056\\
69.82	0.00232504764779125\\
90	-0.0048030270834815\\
100.07	-0.00576760776776173\\
30.24	0.0150070817217082\\
};
\end{axis}
\end{tikzpicture}%

%% file: matlab/tikz_relError_Eureqa.tex
%
%
\definecolor{mycolor1}{rgb}{0.49400,0.18400,0.55600}%
\definecolor{mycolor2}{rgb}{1.00000,0.00000,1.00000}%
\definecolor{mycolor3}{rgb}{0.00000,0.44700,0.74100}%
\definecolor{mycolor4}{rgb}{0.30100,0.74500,0.93300}%
\definecolor{mycolor5}{rgb}{0.46600,0.67400,0.18800}%
\definecolor{mycolor6}{rgb}{0.92900,0.69400,0.12500}%
\definecolor{mycolor7}{rgb}{0.63500,0.07800,0.18400}%
\begin{tikzpicture}[]
\node[] at (4,3) {greatest deviation: \SI[round-mode=figures, round-precision=4]{-0.0181482686682804}{\%}};
\begin{axis}[%
width=\textwidth,
height=1.7in,
xmin=0,
xmax=105,
ymin=-0.025,
ymax=0.032,
ylabel style={font=\color{white!15!black}, text opacity=0},
ylabel={$\text{100}\cdot\text{(}\rho{}_{\text{exp}}\text{)/}\rho{}_{\text{exp}}$ / \si{\%}},
scaled ticks=false,
y tick label style={/pgf/number format/.cd,fixed,precision=2,zerofill},
axis background/.style={fill=white},
every axis title/.style={right,at={(-0.0,1.1)}},
title={b)},
]
\addplot [color=black, forget plot]
  table[row sep=crcr]{%
0	0\\
110	0\\
};
\addplot [color=black, only marks, mark=triangle, mark options={solid, rotate=180, black}, forget plot]
  table[row sep=crcr]{%
5.28	0.00252179189974646\\
9.85	0.00411392124769609\\
15.02	0.00620925407029958\\
20.09	0.00667959392753948\\
30.12	0.00635889985130845\\
49.89	-0.00189797878969907\\
70.07	-0.00142524039374792\\
89.83	0.00251838859009892\\
100.03	0.0126859570047482\\
30.12	0.00348784352933268\\
};
\addplot [color=mycolor1, only marks, mark=+, mark options={solid, mycolor1}, forget plot]
  table[row sep=crcr]{%
4.98	-0.00109861329713161\\
10.1	3.26896417771408e-05\\
15.15	-0.00120922099657125\\
19.96	0.000842615683433162\\
30.15	-0.000799327355961487\\
49.88	-0.00751379199849108\\
69.49	-0.0103784149529267\\
89.98	-0.00556355166834879\\
99.98	0.00260407078178367\\
30.13	0.00211240922038655\\
};
\addplot [color=mycolor1, only marks, mark=triangle, mark options={solid, rotate=270, mycolor1}, forget plot]
table[row sep=crcr]{%
5.04	-0.00711419557765681\\
10.16	-0.00340297256460179\\
15	0.00202432942684614\\
19.84	0.00112798197437204\\
29.94	-0.00067801697898019\\
49.86	-0.00747258143702154\\
69.97	-0.00800081838279702\\
89.75	-0.00233594308341627\\
100	0.00506858199630078\\
30.68	0.000363132306217052\\
};
\addplot [color=mycolor2, only marks, mark=o, mark options={solid, mycolor2}, forget plot]
  table[row sep=crcr]{%
5.08	-0.00195497916808169\\
10.25	0.00169430950199419\\
15.33	0.00434330182034494\\
19.91	0.00426609638138667\\
30.18	0.00269810753939441\\
49.93	-0.00701210936898279\\
70.05	-0.00619849962092677\\
89.99	-0.00269472698211076\\
99.94	0.00611347047711736\\
};
\addplot [color=mycolor3, only marks, mark=asterisk, mark options={solid, mycolor3}, forget plot]
  table[row sep=crcr]{%
5.05	-0.005736203433018\\
10.03	9.4695540569696e-06\\
15.02	0.00184410704809349\\
20.05	0.00355033719263499\\
30.19	0.0026665062352806\\
49.84	-0.00611510488468223\\
69.9	-0.0115743531718742\\
89.94	-0.0087695792102991\\
99.65	-0.00105512531653913\\
30	0.000522482866028933\\
};
\addplot [color=mycolor4, only marks, mark=x, mark options={solid, mycolor4}, forget plot]
  table[row sep=crcr]{%
4.83	-0.00442933270304536\\
10.03	0.00282475502363861\\
15.07	0.00769710767861255\\
20.07	0.00920307344347735\\
29.96	0.00930872947421693\\
50.01	-0.000622278493711737\\
70.18	-0.00504720060737126\\
90.19	-0.00392308643867641\\
100.1	0.000355843339750798\\
30.51	0.00681533834832737\\
};
\addplot [color=mycolor5, only marks, mark=square, mark options={solid, mycolor5}, forget plot]
  table[row sep=crcr]{%
4.94	-0.0150373027721987\\
9.87	-0.00667292665484545\\
15.22	0.00181265327707579\\
19.97	0.00427045576751716\\
30.16	0.00597689562492326\\
49.89	-0.00823853766873054\\
70.02	-0.00777308090299726\\
89.98	-0.00628931013804135\\
100.01	-0.00265409845083329\\
30.21	0.00404339454377473\\
};
\addplot [color=mycolor6, only marks, mark=diamond, mark options={solid, mycolor6}, forget plot]
  table[row sep=crcr]{%
5.04	-0.0181482686682804\\
9.99	-0.00676786843256271\\
14.98	0.0100513754429344\\
19.91	0.0145832678486129\\
30.04	0.0170375889638206\\
49.96	0.00341371165396281\\
69.96	-0.00148454069411413\\
90	0.00138850358095888\\
100.05	0.00698201741415521\\
30.14	0.0103122035225655\\
};
\addplot [color=mycolor7, only marks, mark=triangle, mark options={solid, mycolor7}, forget plot]
  table[row sep=crcr]{%
5	0.0179799484675313\\
9.84	-0.0178238483501917\\
15.1	-0.00777268653660087\\
19.95	-0.00277412648436992\\
29.95	0.00103627637915191\\
49.93	-0.00383909974203834\\
69.82	-0.00798843087821471\\
90	-0.00519837734991236\\
100.07	0.00796300786863378\\
30.24	0.01165828564742\\
};
\end{axis}
\end{tikzpicture}%

%% file: matlab/tikz_relError_fixedExp_5measIsot.tex
%
%
\definecolor{mycolor1}{rgb}{0.49400,0.18400,0.55600}%
\definecolor{mycolor2}{rgb}{1.00000,0.00000,1.00000}%
\definecolor{mycolor3}{rgb}{0.00000,0.44700,0.74100}%
\definecolor{mycolor4}{rgb}{0.30100,0.74500,0.93300}%
\definecolor{mycolor5}{rgb}{0.46600,0.67400,0.18800}%
\definecolor{mycolor6}{rgb}{0.92900,0.69400,0.12500}%
\definecolor{mycolor7}{rgb}{0.63500,0.07800,0.18400}%
\begin{tikzpicture}
\node[] at (4,3) {greatest deviation: \SI[round-mode=figures, round-precision=4]{0.025418936323532}{\%}};
\begin{axis}[%
width=\textwidth,
height=1.7in,
xmin=0,
xmax=105,
ymin=-0.025,
ymax=0.032,
ylabel style={font=\color{white!15!black}, text opacity=0},
ylabel={$\text{100}\cdot\text{(}\rho{}_{\text{exp}}\text{)/}$ / \si{\%}},
scaled ticks=false,
y tick label style={/pgf/number format/.cd,fixed,precision=2,zerofill},
axis background/.style={fill=white},
every axis title/.style={right,at={(-0.0,1.1)}},
title={c)},
]
\addplot [color=black, forget plot]
table[row sep=crcr]{%
	0	0\\
	110	0\\
};
\addplot [color=black, line width=2.0pt, only marks, mark size=2.8pt, mark=triangle, mark options={solid, rotate=180, black}, forget plot]
  table[row sep=crcr]{%
5.28	0.00965865285293209\\
9.85	0.00442886632438272\\
15.02	0.0011939738637934\\
20.09	-0.00131447481012585\\
30.12	-0.00300337414723597\\
49.89	-0.0072491349520708\\
70.07	-0.00585280964318737\\
89.83	-0.00790831979282183\\
100.03	-0.00318688761160411\\
};
\addplot [color=mycolor1, only marks, mark=+, mark options={solid, mycolor1}, forget plot]
  table[row sep=crcr]{%
4.98	0.0193468802761371\\
10.1	0.0125870270445161\\
15.15	0.00625600079243109\\
19.96	0.00562474251352675\\
30.15	0.00382549026127081\\
49.88	0.00546279055724302\\
69.49	0.00830813608179945\\
89.98	0.0102854673395359\\
99.98	0.0139088052659018\\
};
\addplot [color=mycolor2, only marks, mark=o, mark options={solid, mycolor2}, forget plot]
  table[row sep=crcr]{%
5.08	0.0209025524105792\\
10.25	0.0163345639645635\\
15.33	0.0136441765744097\\
19.91	0.0110224291803757\\
30.18	0.00956945941374079\\
49.93	0.0100231172427186\\
70.05	0.0186235109996815\\
89.99	0.0206820331610665\\
99.94	0.025282788491329\\
};
\addplot [color=mycolor3, only marks, mark=asterisk, mark options={solid, mycolor3}, forget plot]
  table[row sep=crcr]{%
5.05	0.0162763304390067\\
10.03	0.0130879000229689\\
15.02	0.00851889574455251\\
20.05	0.00654204302397384\\
30.19	0.00544729040335258\\
49.84	0.00983499704361801\\
69.9	0.0168782953142963\\
89.94	0.0213200730864553\\
99.65	0.025418936323532\\
};
\addplot [color=mycolor4, line width=2.0pt, only marks, mark size=2.8pt, mark=x, mark options={solid, mycolor4}, forget plot]
  table[row sep=crcr]{%
4.83	0.00995153033895133\\
10.03	0.00733576536573969\\
15.07	0.00411528337230729\\
20.07	0.000306744755377484\\
29.96	-0.00221929203585979\\
50.01	0.00143998811584308\\
70.18	0.0123450411489933\\
90.19	0.0162938564740968\\
100.1	0.0161768136900794\\
};
\addplot [color=mycolor5, line width=2.0pt, only marks, mark size=2.8pt, mark=square, mark options={solid, mycolor5}, forget plot]
  table[row sep=crcr]{%
4.94	-0.00629765032346789\\
9.87	-0.00514994218785786\\
15.22	-0.00570148610266011\\
19.97	-0.00958341066051179\\
30.16	-0.0136546506752831\\
49.89	-0.0172305927556864\\
70.02	-0.00204788132540306\\
89.98	0.00048610868481401\\
100.01	-0.00248427726943817\\
};
\addplot [color=mycolor6, line width=2.0pt, only marks, mark size=2.8pt, mark=diamond, mark options={solid, mycolor6}, forget plot]
  table[row sep=crcr]{%
5.04	-0.0131914650829605\\
9.99	-0.00210211879663387\\
14.98	0.0087198119020491\\
19.91	0.00703546523895158\\
30.04	0.00202768262163389\\
49.96	-0.00470228393831424\\
69.96	0.00206292528192543\\
90	0.00100235732062375\\
100.05	-0.00395408429227415\\
};
\addplot [color=mycolor7, line width=2.0pt, only marks, mark size=2.8pt, mark=triangle, mark options={solid, mycolor7}, forget plot]
  table[row sep=crcr]{%
5	0.0104019731043877\\
9.84	-0.00924313814534949\\
15.1	0.00291367609316341\\
19.95	0.00538489301816142\\
29.95	0.00366322849135988\\
49.93	0.0029290986840242\\
69.82	0.00578376989265585\\
90	-0.00322522016925504\\
100.07	-0.00632226927307489\\
};
\end{axis}
\end{tikzpicture}%

%% file: matlab/tikz_relError_freeExp_5measIsot.tex
%
%
\definecolor{mycolor1}{rgb}{0.49400,0.18400,0.55600}%
\definecolor{mycolor2}{rgb}{1.00000,0.00000,1.00000}%
\definecolor{mycolor3}{rgb}{0.00000,0.44700,0.74100}%
\definecolor{mycolor4}{rgb}{0.30100,0.74500,0.93300}%
\definecolor{mycolor5}{rgb}{0.46600,0.67400,0.18800}%
\definecolor{mycolor6}{rgb}{0.92900,0.69400,0.12500}%
\definecolor{mycolor7}{rgb}{0.63500,0.07800,0.18400}%
\begin{tikzpicture}
\node[] at (4,3) {greatest deviation: \SI[round-mode=figures, round-precision=4]{-0.0121445785239107}{\%}};
\begin{axis}[%
width=\textwidth,
height=1.7in,
xmin=0,
xmax=105,
xlabel style={font=\color{white!15!black}},
xlabel={Pressure $p$ / \si{\mega\pascal}},
ymin=-0.025,
ymax=0.032,
ylabel style={font=\color{white!15!black}, text opacity=0},
ylabel={$\text{100}\cdot\text{(}\rho{}_{\text{exp}}\text{)/}\rho{}_{\text{exp}}$ / \si{\%}},
scaled ticks=false,
y tick label style={/pgf/number format/.cd,fixed,precision=2,zerofill},
axis background/.style={fill=white},
every axis title/.style={right,at={(-0.0,1.1)}},
title={d)},
]
\addplot [color=black, forget plot]
table[row sep=crcr]{%
	0	0\\
	110	0\\
};
\addplot [color=black, line width=2.0pt, only marks, mark size=2.8pt, mark=triangle, mark options={solid, rotate=180, black}, forget plot]
table[row sep=crcr]{%
5.28	-0.00173132621307848\\
9.85	-0.00210759518727899\\
15.02	-0.000286723363243653\\
20.09	0.00116336922151741\\
30.12	0.00413729759310806\\
49.89	3.90958617268931e-06\\
70.07	-0.00212230099501913\\
89.83	-0.00404993091689733\\
100.03	0.00361795380691995\\
};
\addplot [color=mycolor1, only marks, mark=+, mark options={solid, mycolor1}, forget plot]
table[row sep=crcr]{%
4.98	0.00301443922543406\\
10.1	0.000480538358000828\\
15.15	-0.00176213795330727\\
19.96	0.000920909760308118\\
30.15	0.00303143766033279\\
49.88	0.00283858460873367\\
69.49	-0.000600029120713872\\
89.98	-0.00191675504216727\\
99.98	0.00290149586006327\\
};
\addplot [color=mycolor2, only marks, mark=o, mark options={solid, mycolor2}, forget plot]
  table[row sep=crcr]{%
5.08	0.00416551968057283\\
10.25	0.00334275004397857\\
15.33	0.00451360718421272\\
19.91	0.00480006058131557\\
30.18	0.00703123640112924\\
49.93	0.00495060350379231\\
70.05	0.00608948445881332\\
89.99	0.00366170208932363\\
99.94	0.00874833133279744\\
};
\addplot [color=mycolor3, only marks, mark=asterisk, mark options={solid, mycolor3}, forget plot]
  table[row sep=crcr]{%
5.05	0.0028367524569152\\
10.03	0.00152746288686933\\
15.02	-0.000201158997185661\\
20.05	0.000609345050652498\\
30.19	0.00285479213750516\\
49.84	0.0042827367061751\\
69.9	0.00226079102523303\\
89.94	-0.00024818264681762\\
99.65	0.00295482296494872\\
};
\addplot [color=mycolor4, line width=2.0pt, only marks, mark size=2.8pt, mark=x, mark options={solid, mycolor4}, forget plot]
table[row sep=crcr]{%
4.83	0.00618616274122544\\
10.03	0.00290145218946768\\
15.07	0.00125979380818726\\
20.07	-0.000403695485014179\\
29.96	0.000395922994845187\\
50.01	0.00245746264075027\\
70.18	0.00503164981899814\\
90.19	0.00186631953833452\\
100.1	0.000551320607381693\\
};
\addplot [color=mycolor5, line width=2.0pt, only marks, mark size=2.8pt, mark=square, mark options={solid, mycolor5}, forget plot]
  table[row sep=crcr]{%
4.94	-0.0020634273141384\\
9.87	-0.00356528694647109\\
15.22	-0.00390215845976537\\
19.97	-0.00653174248093453\\
30.16	-0.00751947854431321\\
49.89	-0.0106857960472717\\
70.02	-0.000720642129765455\\
89.98	-0.00267856894166389\\
100.01	-0.00571602373837988\\
};
\addplot [color=mycolor6, line width=2.0pt, only marks, mark size=2.8pt, mark=diamond, mark options={solid, mycolor6}, forget plot]
table[row sep=crcr]{%
5.04	-0.00724608746070784\\
9.99	0.000968334523008004\\
14.98	0.0111727365186612\\
19.91	0.00966705601114443\\
30.04	0.00620810060839327\\
49.96	0.000638901809916343\\
69.96	0.00571234393063567\\
90	0.00465662566798132\\
100.05	0.00200705046458329\\
};
\addplot [color=mycolor7, line width=2.0pt, only marks, mark size=2.8pt, mark=triangle, mark options={solid, mycolor7}, forget plot]
table[row sep=crcr]{%
5	0.00876323779453272\\
9.84	-0.0121445785239107\\
15.1	-0.00047121330051328\\
19.95	0.00126925957299683\\
29.95	-0.00159419597403472\\
49.93	-0.00301470925442662\\
69.82	0.000914442818652949\\
90	-0.0026810266841061\\
100.07	-0.00022067589163415\\
};
\end{axis}
\end{tikzpicture}%

%% file: matlab/tikz_relError_fixedExp_9measIsot_PG.tex
%
%
\definecolor{mycolor1}{rgb}{0.49400,0.18400,0.55600}%
\definecolor{mycolor2}{rgb}{1.00000,0.00000,1.00000}%
\definecolor{mycolor3}{rgb}{0.00000,0.44700,0.74100}%
\definecolor{mycolor4}{rgb}{0.30100,0.74500,0.93300}%
\definecolor{mycolor5}{rgb}{0.46600,0.67400,0.18800}%
\definecolor{mycolor6}{rgb}{0.92900,0.69400,0.12500}%
\definecolor{mycolor8}{rgb}{0.87000,0.49000,0.00000}%
\definecolor{mycolor7}{rgb}{0.63500,0.07800,0.18400}%
\begin{tikzpicture}
\node[] at (4,3) {greatest deviation: \SI[round-mode=figures, round-precision=4]{0.014343}{\%}};
\begin{axis}[%
	width=\textwidth,
	height=1.7in,
	xmin=0,
	xmax=105,
	ymin=-0.025,
	ymax=0.032,
	ylabel style={font=\color{white!15!black}, text opacity=0},
	ylabel={$\text{100}\cdot\text{(}\rho{}_{\text{exp}}\text{)/}\rho{}_{\text{exp}}$ / \si{\%}},
	scaled ticks=false,
	y tick label style={/pgf/number format/.cd,fixed,precision=2,zerofill},
	axis background/.style={fill=white},
	every axis title/.style={right,at={(-0.0,1.1)}},
	title={a)},
	]
	\addplot [color=black, forget plot]
	table[row sep=crcr]{%
		0	0\\
		110	0\\
	};
\addplot [color=black, only marks, mark=triangle, mark options={solid, rotate=180, black}, forget plot]
  table[row sep=crcr]{%
5.24	0.0142830643706315\\
10.29	0.0125784529542793\\
15.27	0.00730258495542932\\
20.36	0.00305633948248704\\
30.58	0.00187415882466202\\
50.68	-0.00856930666703051\\
71	-0.00813928325332339\\
91.4	-0.00172489775529738\\
};
\addplot [color=mycolor1, only marks, mark=+, mark options={solid, mycolor1}, forget plot]
  table[row sep=crcr]{%
5.19	0.00785559154712457\\
10.06	0.00381008005098902\\
15.1	0.00124684257752628\\
20.2	-0.000471574037562165\\
30.53	-0.00506359334539826\\
50.55	-0.0120263412240142\\
70.99	-0.0119109750002805\\
91.27	-0.00209924459308919\\
};
\addplot [color=mycolor2, only marks, mark=o, mark options={solid, mycolor2}, forget plot]
  table[row sep=crcr]{%
5.18	0.00206887907603291\\
10.1	0.00134875717777506\\
15.21	-9.21984823872027e-05\\
20.19	-0.000833681671867342\\
30.24	-0.0045037941995824\\
50.55	-0.00609344070831764\\
70.76	-0.00510030299100942\\
90.99	0.00426324062098526\\
};
\addplot [color=mycolor3, only marks, mark=asterisk, mark options={solid, mycolor3}, forget plot]
  table[row sep=crcr]{%
5.09	0.00331386372601318\\
10.07	0.00201314960472545\\
15.24	0.00108114306511087\\
20.17	0.000221768450793492\\
30.3	-0.00198816199384444\\
50.59	-0.00157202547723339\\
70.93	0.000557381939732113\\
91.05	0.0110858886392603\\
};
\addplot [color=mycolor4, only marks, mark=x, mark options={solid, mycolor4}, forget plot]
  table[row sep=crcr]{%
5.12	-0.0019845498839091\\
10.07	-0.00546733720162238\\
15.11	-0.0057137100890692\\
20.29	-0.00631827834290584\\
30.29	-0.00701306974474148\\
50.71	-0.00453563664050302\\
71.03	0.000817225914151227\\
91.24	0.0111158390116439\\
};
\addplot [color=mycolor5, only marks, mark=square, mark options={solid, mycolor5}, forget plot]
  table[row sep=crcr]{%
5.26	-0.00239837131876834\\
10.05	-0.00546871159507558\\
15.12	-0.0058325382742878\\
20.25	-0.00357319552216\\
30.4	-0.00269205525055795\\
50.61	1.31759293196957e-05\\
70.91	0.00581990526731815\\
91.13	0.0143433033364009\\
};
\addplot [color=mycolor6, only marks, mark=diamond, mark options={solid, mycolor6}, forget plot]
  table[row sep=crcr]{%
5.01	0.00551028875491191\\
10.1	0.00536392966342872\\
15.28	0.00153238743751721\\
20.21	0.00212758521086632\\
30.36	0.00128740339461787\\
50.61	0.00680215995182814\\
70.75	0.00865637445012394\\
91.06	0.0117985315070708\\
};
\addplot [color=mycolor8, only marks, mark options={solid, rotate=90, mycolor8}, forget plot]
  table[row sep=crcr]{%
10.32	-0.00679814368214413\\
15.33	-0.00471104268184032\\
20.36	-0.0106365256280332\\
30.53	-0.00477879947224539\\
50.61	-0.00669820980042113\\
70.99	-0.0020441590310357\\
91.19	-0.0119396221770702\\
50.86	-0.00295440814416607\\
};
\addplot [color=mycolor7, only marks, mark=triangle, mark options={solid, mycolor7}, forget plot]
  table[row sep=crcr]{%
5.08	-0.00510390433015737\\
10.09	0.0018440105361273\\
15.19	0.00601071709784676\\
20.23	0.00966069098176822\\
30.43	0.0087106944195064\\
50.73	0.00689154453601288\\
70.89	0.00103345532391349\\
90.95	-0.0102458270984266\\
};
\end{axis}
\end{tikzpicture}%

%% file: matlab/tikz_relError_fixedExp_5measIsot_PG.tex
%
%
\definecolor{mycolor1}{rgb}{0.49400,0.18400,0.55600}%
\definecolor{mycolor2}{rgb}{1.00000,0.00000,1.00000}%
\definecolor{mycolor3}{rgb}{0.00000,0.44700,0.74100}%
\definecolor{mycolor4}{rgb}{0.30100,0.74500,0.93300}%
\definecolor{mycolor5}{rgb}{0.46600,0.67400,0.18800}%
\definecolor{mycolor6}{rgb}{0.92900,0.69400,0.12500}%
\definecolor{mycolor8}{rgb}{0.87000,0.49000,0.00000}%
\definecolor{mycolor7}{rgb}{0.63500,0.07800,0.18400}%
\begin{tikzpicture}
\node[] at (4,3) {greatest deviation: \SI[round-mode=figures, round-precision=4]{0.015707}{\%}};
\begin{axis}[%
	width=\textwidth,
	height=1.7in,
	xmin=0,
	xmax=105,
	ymin=-0.025,
	ymax=0.032,
	ylabel style={font=\color{white!15!black}, text opacity=0},
	ylabel={$\text{100}\cdot\text{(}\rho{}_{\text{exp}}\text{)/}\rho{}_{\text{exp}}$ / \si{\%}},
	scaled ticks=false,
	y tick label style={/pgf/number format/.cd,fixed,precision=2,zerofill},
	axis background/.style={fill=white},
	every axis title/.style={right,at={(-0.0,1.1)}},
	title={b)},
	]
	\addplot [color=black, forget plot]
	table[row sep=crcr]{%
		0	0\\
		110	0\\
	};
\addplot [color=black, line width=2.0pt, only marks, mark size=2.8pt, mark=triangle, mark options={solid, rotate=180, black}, forget plot]
  table[row sep=crcr]{%
5.24	0.0117334672993494\\
10.29	0.00945928858593472\\
15.27	0.0037211050804946\\
20.36	-0.000861983293592159\\
30.58	-0.00228739528609652\\
50.68	-0.0117495362374273\\
71	-0.00893499531142492\\
91.4	0.000799859861763304\\
};
\addplot [color=mycolor1, only marks, mark=+, mark options={solid, mycolor1}, forget plot]
  table[row sep=crcr]{%
5.19	0.00594405785672053\\
10.06	0.00122906176899962\\
15.1	-0.00195744052124231\\
20.2	-0.004176140577811\\
30.53	-0.00931226146107955\\
50.55	-0.0157070107886611\\
70.99	-0.013415293964512\\
91.27	-0.000403370110456286\\
};
\addplot [color=mycolor2, only marks, mark=o, mark options={solid, mycolor2}, forget plot]
  table[row sep=crcr]{%
5.18	0.00081783995418686\\
10.1	-0.00065731567801303\\
15.21	-0.002852891874346\\
20.19	-0.00421817622632205\\
30.24	-0.00869247082511674\\
50.55	-0.0101082952183292\\
70.76	-0.00716307073742516\\
90.99	0.00531412009620751\\
};
\addplot [color=mycolor3, only marks, mark=asterisk, mark options={solid, mycolor3}, forget plot]
  table[row sep=crcr]{%
5.09	0.00241130295315671\\
10.07	0.00032227101109702\\
15.24	-0.0014268965885049\\
20.17	-0.00296534509048045\\
30.3	-0.00611193501037758\\
50.59	-0.00570340347029539\\
70.93	-0.00169172802124884\\
91.05	0.0119166598383521\\
};
\addplot [color=mycolor4, only marks, mark=x, mark options={solid, mycolor4}, forget plot]
  table[row sep=crcr]{%
5.12	-0.00190681972866587\\
10.07	-0.00617234756199961\\
15.11	-0.00732659291669149\\
20.29	-0.00881166307251159\\
30.29	-0.0107797749514881\\
50.71	-0.00884654316712209\\
71.03	-0.00184365231410373\\
91.24	0.0115361244484064\\
};
\addplot [color=mycolor5, line width=2.0pt, only marks, mark size=2.8pt, mark=square, mark options={solid, mycolor5}, forget plot]
  table[row sep=crcr]{%
5.26	-0.00132904637403798\\
10.05	-0.00494292888719995\\
15.12	-0.00621803091278396\\
20.25	-0.00495274657076537\\
30.4	-0.00572723403817979\\
50.61	-0.00416503379062598\\
70.91	0.00306406157425985\\
91.13	0.0147682792651242\\
};
\addplot [color=mycolor6, line width=2.0pt, only marks, mark size=2.8pt, mark=diamond, mark options={solid, mycolor6}, forget plot]
  table[row sep=crcr]{%
5.01	0.00681690853257927\\
10.1	0.00670433716170786\\
15.28	0.00217762522740317\\
20.21	0.00185352504728901\\
30.36	-0.000816520122939076\\
50.61	0.00314248417775465\\
70.75	0.00634689327939005\\
91.06	0.0129698890359071\\
};
\addplot [color=mycolor8, line width=2.0pt, only marks, mark size=2.8pt, mark=triangle, mark options={solid, rotate=90, mycolor8}, forget plot]
  table[row sep=crcr]{%
10.32	-0.00543502736674649\\
15.33	-0.0034830221362962\\
20.36	-0.0100869014089236\\
30.53	-0.00594699513835821\\
50.61	-0.00950055031381843\\
70.99	-0.0032823548435423\\
91.19	-0.00914930945017606\\
50.86	-0.00575633006604434\\
};
\addplot [color=mycolor7, line width=2.0pt, only marks, mark size=2.8pt, mark=triangle, mark options={solid, mycolor7}, forget plot]
  table[row sep=crcr]{%
5.08	-0.00772619841539481\\
10.09	0.00194844481133342\\
15.19	0.00697399717332026\\
20.23	0.010500169452916\\
30.43	0.0083144234448292\\
50.73	0.00515539852464995\\
70.89	0.00131837513270693\\
90.95	-0.00515693512201318\\
};
\end{axis}
\end{tikzpicture}%

%% file: matlab/tikz_relError_freeExp_5measIsot_PG.tex
%
%
\definecolor{mycolor1}{rgb}{0.49400,0.18400,0.55600}%
\definecolor{mycolor2}{rgb}{1.00000,0.00000,1.00000}%
\definecolor{mycolor3}{rgb}{0.00000,0.44700,0.74100}%
\definecolor{mycolor4}{rgb}{0.30100,0.74500,0.93300}%
\definecolor{mycolor5}{rgb}{0.46600,0.67400,0.18800}%
\definecolor{mycolor6}{rgb}{0.92900,0.69400,0.12500}%
\definecolor{mycolor8}{rgb}{0.87000,0.49000,0.00000}%
\definecolor{mycolor7}{rgb}{0.63500,0.07800,0.18400}%
\begin{tikzpicture}
\node[] at (4,3) {greatest deviation: \SI[round-mode=figures, round-precision=4]{0.043053}{\%}};
\begin{axis}[%
	width=\textwidth,
	height=1.7in,
	xmin=0,
	xmax=105,
	ymin=-0.025,
	ymax=0.044,
	ylabel style={font=\color{white!15!black}, text opacity=0},
	ylabel={$\text{100}\cdot\text{(}\rho{}_{\text{exp}}\text{)/}\rho{}_{\text{exp}}$ / \si{\%}},
	scaled ticks=false,
	y tick label style={/pgf/number format/.cd,fixed,precision=2,zerofill},
	axis background/.style={fill=white},
	every axis title/.style={right,at={(-0.0,1.1)}},
	title={c)},
	]
	\addplot [color=black, forget plot]
	table[row sep=crcr]{%
		0	0\\
		110	0\\
	};
\addplot [color=black, line width=2.0pt, only marks, mark size=2.8pt, mark=triangle, mark options={solid, rotate=180, black}, forget plot]
  table[row sep=crcr]{%
5.24	-0.000483138154477233\\
10.29	0.00134971199727446\\
15.27	-0.000678842100090711\\
20.36	-0.00167750926238838\\
30.58	0.00236171895129242\\
50.68	-0.00128784414265028\\
71	0.00047475823851001\\
91.4	-3.50290612785691e-05\\
};
\addplot [color=mycolor1, only marks, mark=+, mark options={solid, mycolor1}, forget plot]
  table[row sep=crcr]{%
5.19	0.0330040776160221\\
10.06	0.0315755009282874\\
15.1	0.0313779757958467\\
20.2	0.0317183284698622\\
30.53	0.0302785134395573\\
50.55	0.0262443223617443\\
70.99	0.0247509736080326\\
91.27	0.0263423212872055\\
};
\addplot [color=mycolor2, only marks, mark=o, mark options={solid, mycolor2}, forget plot]
  table[row sep=crcr]{%
5.18	0.0400871212393543\\
10.1	0.0414760826335114\\
15.21	0.0418374392774674\\
20.19	0.042424839822504\\
30.24	0.0403147492781963\\
50.55	0.0386687994220722\\
70.76	0.0359058524326379\\
90.99	0.0362605151283888\\
};
\addplot [color=mycolor3, only marks, mark=asterisk, mark options={solid, mycolor3}, forget plot]
  table[row sep=crcr]{%
5.09	0.0415783640878037\\
10.07	0.042197726965092\\
15.24	0.0428525528440704\\
20.17	0.043053322643765\\
30.3	0.041811383227179\\
50.59	0.0409243087501792\\
70.93	0.0383926814570786\\
91.05	0.0396651140806964\\
};
\addplot [color=mycolor4, only marks, mark=x, mark options={solid, mycolor4}, forget plot]
  table[row sep=crcr]{%
5.12	0.0235413788811101\\
10.07	0.0215342423810513\\
15.11	0.0224210509479959\\
20.29	0.0224301632132625\\
30.29	0.021472359412735\\
50.71	0.0199346763021443\\
71.03	0.0189370235223552\\
91.24	0.0199786700101642\\
};
\addplot [color=mycolor5, line width=2.0pt, only marks, mark size=2.8pt, mark=square, mark options={solid, mycolor5}, forget plot]
  table[row sep=crcr]{%
5.26	0.000121531208013506\\
10.05	-0.00185548647140026\\
15.12	-0.0011917407332645\\
20.25	0.00158119488465754\\
30.4	0.00180807864802395\\
50.61	-0.000580603823344626\\
70.91	-0.00106334313058292\\
91.13	0.000506002501211773\\
};
\addplot [color=mycolor6, line width=2.0pt, only marks, mark size=2.8pt, mark=diamond, mark options={solid, mycolor6}, forget plot]
  table[row sep=crcr]{%
5.01	0.00164958883514652\\
10.1	0.00183883644402007\\
15.28	-0.00107150459819872\\
20.21	0.000167352918763415\\
30.36	-0.00081586368709437\\
50.61	0.000786888813672834\\
70.75	-0.00108182909860675\\
91.06	-2.95780753698279e-05\\
};
\addplot [color=mycolor8, line width=2.0pt, only marks, mark size=2.8pt, mark=triangle, mark options={solid, rotate=90, mycolor8}, forget plot]
  table[row sep=crcr]{%
10.32	-0.00106686528663526\\
15.33	0.00094431877075403\\
20.36	-0.00460757145799918\\
30.53	0.00146415229551065\\
50.61	-0.00201141472961087\\
70.99	0.00345680675997039\\
91.19	-0.00113826194556227\\
50.86	0.00171882631374045\\
};
\addplot [color=mycolor7, line width=2.0pt, only marks, mark size=2.8pt, mark=triangle, mark options={solid, mycolor7}, forget plot]
  table[row sep=crcr]{%
5.08	-0.000307777784784061\\
10.09	-0.000694752850092173\\
15.19	0.0002719395128723\\
20.23	0.00258665261044144\\
30.43	0.000523986568334422\\
50.73	-0.0014673526331\\
70.89	-0.00153033566455455\\
90.95	0.00090224230299205\\
};
\end{axis}
\end{tikzpicture}%

%% file: matlab/tikz_extrapol2VLsat_pressure.tex
%
%
\definecolor{mycolor1}{rgb}{0.00000,0.44700,0.74100}%
\definecolor{mycolor2}{rgb}{0.85000,0.32500,0.09800}%
\definecolor{mycolor3}{rgb}{0.92900,0.69400,0.12500}%
\definecolor{mycolor4}{rgb}{0.49400,0.18400,0.55600}%
\definecolor{mycolor5}{rgb}{0.46600,0.67400,0.18800}%
\definecolor{mycolor6}{rgb}{0.30100,0.74500,0.93300}%
\definecolor{mycolor7}{rgb}{0.63500,0.07800,0.18400}%
\begin{tikzpicture}

\begin{axis}[%
width=\textwidth,
height=1.7in,
xmin=0,
xmax=10.6,
xlabel style={font=\color{white!15!black}},
xlabel={Pressure \(p\) / \si{MPa}},
ymin=300,
ymax=1200,
ylabel style={font=\color{white!15!black}, text opacity=0},
ylabel={Density / \si{\kilo\gram\per\meter\cubed}},
scaled ticks=false,
y tick label style={/pgf/number format/.cd,fixed,precision=0,zerofill},
axis background/.style={fill=white},
every axis title/.style={right,at={(-0.0,1.1)}},
title={a)},
legend style={legend cell align=left, align=left, draw=white!15!black, font=\footnotesize}
]
\addplot [color=mycolor2, line width=1]
  table[row sep=crcr]{%
7.102e-07	1127.94011609211\\
1.2308e-06	1124.42952887082\\
2.0858e-06	1120.92818236383\\
3.4604e-06	1117.43353946207\\
5.6273e-06	1113.94299757413\\
8.9792e-06	1110.45388089963\\
1.4073e-05	1106.96343400392\\
2.1683e-05	1103.46881640518\\
3.2874e-05	1099.96709796056\\
4.9082e-05	1096.45525488038\\
7.2221e-05	1092.93016625158\\
0.0001048	1089.38861096663\\
0.0001501	1085.82726500681\\
0.00021228	1082.2426989636\\
0.00029665	1078.63137586783\\
0.00040984	1074.98964917091\\
0.00056009	1071.31376096603\\
0.0007575	1067.59984036031\\
0.0010143	1063.84390202139\\
0.0013454	1060.04184505623\\
0.0017685	1056.18945182532\\
0.0023044	1052.28238704794\\
0.0029777	1048.31619737409\\
0.0038173	1044.28631110719\\
0.0048564	1040.18803785063\\
0.0061331	1036.0165688465\\
0.0076913	1031.76697793624\\
0.0095806	1027.43422216575\\
0.011857	1023.01314342167\\
0.014583	1018.49847005235\\
0.017829	1013.88482022397\\
0.021673	1009.16670438224\\
0.026202	1004.33853002842\\
0.031509	999.394602439823\\
0.037698	994.329134716977\\
0.044882	989.136251320456\\
0.053184	983.809995684085\\
0.062736	978.344336162323\\
0.073681	972.73317689134\\
0.086172	966.970365664705\\
0.10037	961.04970057141\\
0.11646	954.964968310054\\
0.13462	948.709904778011\\
0.15504	942.278237244036\\
0.17795	935.663756687476\\
0.20355	928.86019628983\\
0.23208	921.861385675351\\
0.26378	914.661188081705\\
0.2989	907.253537433176\\
0.33773	899.632536013316\\
0.38052	891.792270455668\\
0.42758	883.727090251701\\
0.47921	875.431447708846\\
0.53572	866.899968977429\\
0.59744	858.127508199702\\
0.6647	849.109105792611\\
0.73787	839.840147637516\\
0.8173	830.316180146709\\
0.90338	820.533144778029\\
0.99648	810.487172653006\\
1.097	800.174792838059\\
1.2054	789.593099827568\\
1.3221	778.739345324598\\
1.4476	767.611487977766\\
1.5822	756.206859757936\\
1.7265	744.524584434482\\
1.881	732.563689556067\\
2.0462	720.323563753588\\
2.2226	707.803943416564\\
2.4109	695.006081009199\\
2.6115	681.929848221022\\
2.8251	668.57721133838\\
3.0523	654.949879742871\\
3.2939	641.051169333919\\
3.5505	626.883403391181\\
3.8228	612.449780839887\\
4.1118	597.755827923958\\
4.4182	582.805188256601\\
4.743	567.603773611009\\
5.0873	552.158380628597\\
5.4521	536.475107152007\\
5.8387	520.562304116017\\
6.2484	504.428239638558\\
6.6829	488.084167146817\\
7.1441	471.542715525261\\
7.6341	454.81775022637\\
8.1557	437.928335790681\\
8.7122	420.896951960857\\
9.3083	403.756715703408\\
};

\addplot [color=mycolor3, line width=1]
  table[row sep=crcr]{%
7.102e-07	1128.34093798069\\
1.2308e-06	1124.59472179173\\
2.0858e-06	1120.92526779919\\
3.4604e-06	1117.31590382811\\
5.6273e-06	1113.75233949619\\
8.9792e-06	1110.22225558433\\
1.4073e-05	1106.71497554328\\
2.1683e-05	1103.22120059084\\
3.2874e-05	1099.73279449614\\
4.9082e-05	1096.24260750983\\
7.2221e-05	1092.7443313918\\
0.0001048	1089.23237931516\\
0.0001501	1085.70178583685\\
0.00021228	1082.14812308647\\
0.00029665	1078.56743029628\\
0.00040984	1074.9561541695\\
0.00056009	1071.31109829513\\
0.0007575	1067.6293800032\\
0.0010143	1063.90839345211\\
0.0013454	1060.14577808511\\
0.0017685	1056.33939125073\\
0.0023044	1052.4872846421\\
0.0029777	1048.58768412335\\
0.0038173	1044.63897216099\\
0.0048564	1040.63967227591\\
0.0061331	1036.58843583475\\
0.0076913	1032.48403077304\\
0.0095806	1028.32533120676\\
0.011857	1024.11130884174\\
0.014583	1019.8410251445\\
0.017829	1015.51362539433\\
0.021673	1011.1283325067\\
0.026202	1006.68444281301\\
0.031509	1002.18131938134\\
0.037698	997.61839168459\\
0.044882	992.995150866031\\
0.053184	988.311147646667\\
0.062736	983.56598914167\\
0.073681	978.759338413668\\
0.086172	973.890912052342\\
0.10037	968.960477102283\\
0.11646	963.967863755303\\
0.13462	958.912938152844\\
0.15504	953.795617511331\\
0.17795	948.615895817611\\
0.20355	943.373780189189\\
0.23208	938.069354578996\\
0.26378	932.70274333015\\
0.2989	927.27412045961\\
0.33773	921.783740636708\\
0.38052	916.231859199132\\
0.42758	910.618833914153\\
0.47921	904.945054799318\\
0.53572	899.21096482314\\
0.59744	893.41707182878\\
0.6647	887.563927427135\\
0.73787	881.652173302279\\
0.8173	875.682474003448\\
0.90338	869.655586977603\\
0.99648	863.57229285065\\
1.097	857.433455449645\\
1.2054	851.240062759389\\
1.3221	844.993103785711\\
1.4476	838.693721426594\\
1.5822	832.342846786094\\
1.7265	825.941855242609\\
1.881	819.492046870104\\
2.0462	812.994774801886\\
2.2226	806.451444285629\\
2.4109	799.863800136277\\
2.6115	793.233214904628\\
2.8251	786.561546589682\\
3.0523	779.850566395844\\
3.2939	773.10240453134\\
3.5505	766.318949742417\\
3.8228	759.502296788553\\
4.1118	752.655072076261\\
4.4182	745.779495422578\\
4.743	738.878323889938\\
5.0873	731.954547962787\\
5.4521	725.011062991454\\
5.8387	718.051329021046\\
6.2484	711.078881486057\\
6.6829	704.098013053921\\
7.1441	697.113452598118\\
7.6341	690.13037149827\\
8.1557	683.15527324379\\
8.7122	676.195674659499\\
9.3083	669.261742146561\\
};

\addplot [color=mycolor4]
  table[row sep=crcr]{%
7.102e-07	1128.04173090871\\
1.2308e-06	1124.53729779221\\
2.0858e-06	1121.04177803525\\
3.4604e-06	1117.55253746682\\
5.6273e-06	1114.06687825694\\
8.9792e-06	1110.58203147722\\
1.4073e-05	1107.09515101048\\
2.1683e-05	1103.60330851156\\
3.2874e-05	1100.10348919952\\
4.9082e-05	1096.59258830522\\
7.2221e-05	1093.06740805142\\
0.0001048	1089.52465505872\\
0.0001501	1085.96093812355\\
0.00021228	1082.37276625142\\
0.00029665	1078.75654701163\\
0.00040984	1075.1085850594\\
0.00056009	1071.42508091078\\
0.0007575	1067.70212988287\\
0.0010143	1063.93572122214\\
0.0013454	1060.12173757718\\
0.0017685	1056.25595442644\\
0.0023044	1052.33403981169\\
0.0029777	1048.35155455316\\
0.0038173	1044.30395263813\\
0.0048564	1040.18658155912\\
0.0061331	1035.99468336454\\
0.0076913	1031.72339635921\\
0.0095806	1027.36775648595\\
0.011857	1022.92269976936\\
0.014583	1018.38306478461\\
0.017829	1013.74359690643\\
0.021673	1008.9989517136\\
0.026202	1004.14370076768\\
0.031509	999.172333360356\\
0.037698	994.079267711567\\
0.044882	988.858855683557\\
0.053184	983.50539167087\\
0.062736	978.013119871713\\
0.073681	972.376246620179\\
0.086172	966.588949769746\\
0.10037	960.645386871652\\
0.11646	954.539735852063\\
0.13462	948.266156280529\\
0.15504	941.818834067924\\
0.17795	935.192057607898\\
0.20355	928.380094690159\\
0.23208	921.377352970715\\
0.26378	914.178317358132\\
0.2989	906.777590237307\\
0.33773	899.169994829046\\
0.38052	891.350386841455\\
0.42758	883.313945457737\\
0.47921	875.056010021338\\
0.53572	866.572156260425\\
0.59744	857.858255462856\\
0.6647	848.910433882785\\
0.73787	839.725242052123\\
0.8173	830.299464815117\\
0.90338	820.630369943951\\
0.99648	810.715496624118\\
1.097	800.552877387002\\
1.2054	790.141218744814\\
1.3221	779.479476462025\\
1.4476	768.567438677264\\
1.5822	757.404323791622\\
1.7265	745.991336810168\\
1.881	734.329684062283\\
2.0462	722.421060336683\\
2.2226	710.267640221065\\
2.4109	697.873325420617\\
2.6115	685.240676131769\\
2.8251	672.374608773125\\
3.0523	659.279906176638\\
3.2939	645.963211485537\\
3.5505	632.430265121065\\
3.8228	618.687905076088\\
4.1118	604.745637276805\\
4.4182	590.611138977265\\
4.743	576.294718180628\\
5.0873	561.807851045746\\
5.4521	547.161491447943\\
5.8387	532.369268126606\\
6.2484	517.44498942344\\
6.6829	502.405986634473\\
7.1441	487.27141212138\\
7.6341	472.062124596731\\
8.1557	456.8050518169\\
8.7122	441.531330568304\\
9.3083	426.284302366029\\
};

\addplot [color=mycolor6]
  table[row sep=crcr]{%
7.102e-07	1128.03415551651\\
1.2308e-06	1124.43816584712\\
2.0858e-06	1120.88027895629\\
3.4604e-06	1117.35284021267\\
5.6273e-06	1113.84883593758\\
8.9792e-06	1110.36177318078\\
1.4073e-05	1106.88557911049\\
2.1683e-05	1103.4145161557\\
3.2874e-05	1099.94310990747\\
4.9082e-05	1096.46608742635\\
7.2221e-05	1092.97832410962\\
0.0001048	1089.47479763774\\
0.0001501	1085.95054784164\\
0.00021228	1082.40064148467\\
0.00029665	1078.82014129639\\
0.00040984	1075.20407851737\\
0.00056009	1071.54742855158\\
0.0007575	1067.8450892444\\
0.0010143	1064.09186148156\\
0.0013454	1060.28243199097\\
0.0017685	1056.41135776057\\
0.0023044	1052.47305222964\\
0.0029777	1048.46177328911\\
0.0038173	1044.37161269634\\
0.0048564	1040.19648660277\\
0.0061331	1035.93012788709\\
0.0076913	1031.56608023313\\
0.0095806	1027.09769297105\\
0.011857	1022.51811808364\\
0.014583	1017.82030835979\\
0.017829	1012.99701861763\\
0.021673	1008.04080729423\\
0.026202	1002.94404196489\\
0.031509	997.698901669845\\
0.037698	992.297391459656\\
0.044882	986.731351328945\\
0.053184	980.992471695927\\
0.062736	975.072308781911\\
0.073681	968.962308233964\\
0.086172	962.653827303505\\
0.10037	956.13815735221\\
0.11646	949.406592856394\\
0.13462	942.450395789105\\
0.15504	935.260873406576\\
0.17795	927.829509627593\\
0.20355	920.147812242201\\
0.23208	912.207569992662\\
0.26378	904.000791330631\\
0.2989	895.519794230256\\
0.33773	886.757403540805\\
0.38052	877.706688849178\\
0.42758	868.361465502643\\
0.47921	858.716076369288\\
0.53572	848.765555784231\\
0.59744	838.50577471575\\
0.6647	827.933417075988\\
0.73787	817.046323937661\\
0.8173	805.843207627781\\
0.90338	794.324142762506\\
0.99648	782.490231065288\\
1.097	770.344051996607\\
1.2054	757.890062596514\\
1.3221	745.13384136748\\
1.4476	732.083247254171\\
1.5822	718.745707127865\\
1.7265	705.13326327056\\
1.881	691.258726703261\\
2.0462	677.13663534075\\
2.2226	662.783239895281\\
2.4109	648.219099132657\\
2.6115	633.462720384972\\
2.8251	618.538209626455\\
3.0523	603.470071667987\\
3.2939	588.287389021595\\
3.5505	573.017876639965\\
3.8228	557.692051014935\\
4.1118	542.346595506983\\
4.4182	527.014493530713\\
4.743	511.73458288162\\
5.0873	496.548250424877\\
5.4521	481.495517989425\\
5.8387	466.621867672791\\
6.2484	451.972498630391\\
6.6829	437.599420827301\\
7.1441	423.557333129357\\
7.6341	409.902925161799\\
8.1557	396.703929025613\\
8.7122	384.034231661629\\
9.3083	371.990266672034\\
};

\addplot [color=mycolor1, line width=1.0pt]
table[row sep=crcr]{%
	7.102e-07	1129.6\\
	1.2308e-06	1126\\
	2.0858e-06	1122.5\\
	3.4604e-06	1119\\
	5.6273e-06	1115.5\\
	8.9792e-06	1112\\
	1.4073e-05	1108.5\\
	2.1683e-05	1105\\
	3.2874e-05	1101.5\\
	4.9082e-05	1097.9\\
	7.2221e-05	1094.4\\
	0.0001048	1090.8\\
	0.0001501	1087.2\\
	0.00021228	1083.6\\
	0.00029665	1080\\
	0.00040984	1076.4\\
	0.00056009	1072.7\\
	0.0007575	1069.1\\
	0.0010143	1065.4\\
	0.0013454	1061.6\\
	0.0017685	1057.9\\
	0.0023044	1054.1\\
	0.0029777	1050.3\\
	0.0038173	1046.5\\
	0.0048564	1042.6\\
	0.0061331	1038.7\\
	0.0076913	1034.8\\
	0.0095806	1030.8\\
	0.011857	1026.8\\
	0.014583	1022.8\\
	0.017829	1018.7\\
	0.021673	1014.6\\
	0.026202	1010.5\\
	0.031509	1006.3\\
	0.037698	1002\\
	0.044882	997.75\\
	0.053184	993.42\\
	0.062736	989.04\\
	0.073681	984.61\\
	0.086172	980.12\\
	0.10037	975.59\\
	0.11646	970.99\\
	0.13462	966.34\\
	0.15504	961.63\\
	0.17795	956.85\\
	0.20355	952.02\\
	0.23208	947.11\\
	0.26378	942.13\\
	0.2989	937.08\\
	0.33773	931.96\\
	0.38052	926.75\\
	0.42758	921.47\\
	0.47921	916.09\\
	0.53572	910.63\\
	0.59744	905.08\\
	0.6647	899.43\\
	0.73787	893.68\\
	0.8173	887.82\\
	0.90338	881.85\\
	0.99648	875.76\\
	1.097	869.55\\
	1.2054	863.21\\
	1.3221	856.73\\
	1.4476	850.11\\
	1.5822	843.33\\
	1.7265	836.39\\
	1.881	829.28\\
	2.0462	821.98\\
	2.2226	814.48\\
	2.4109	806.77\\
	2.6115	798.83\\
	2.8251	790.64\\
	3.0523	782.18\\
	3.2939	773.43\\
	3.5505	764.34\\
	3.8228	754.9\\
	4.1118	745.05\\
	4.4182	734.75\\
	4.743	723.93\\
	5.0873	712.52\\
	5.4521	700.43\\
	5.8387	687.52\\
	6.2484	673.64\\
	6.6829	658.57\\
	7.1441	641.99\\
	7.6341	623.45\\
	8.1557	602.22\\
	8.7122	577.04\\
	9.3083	545.33\\
};

\addplot [color=black,  line width=1.0pt, only marks, mark size=2.8pt, mark=x]
table[row sep=crcr]{%
	10.509	364.96\\
};
\end{axis}
\end{tikzpicture}%

%% file: matlab/tikz_extrapol2VLsat_temperature.tex
%
%
\definecolor{mycolor1}{rgb}{0.00000,0.44700,0.74100}%
\definecolor{mycolor2}{rgb}{0.85000,0.32500,0.09800}%
\definecolor{mycolor3}{rgb}{0.92900,0.69400,0.12500}%
\definecolor{mycolor4}{rgb}{0.49400,0.18400,0.55600}%
\definecolor{mycolor5}{rgb}{0.46600,0.67400,0.18800}%
\definecolor{mycolor6}{rgb}{0.30100,0.74500,0.93300}%
\definecolor{mycolor7}{rgb}{0.63500,0.07800,0.18400}%
\begin{tikzpicture}

\begin{axis}[%
width=\textwidth,
height=1.7in,
xmin=250,
xmax=750,
xlabel style={font=\color{white!15!black}},
xlabel={Temperature \(T\) / K},
ymin=300,
ymax=1200,
ylabel style={font=\color{white!15!black}, text opacity=0},
ylabel={Density / \si{\kilo\gram\per\meter\cubed}},
scaled ticks=false,
y tick label style={/pgf/number format/.cd,fixed,precision=0,zerofill},
axis background/.style={fill=white},
every axis title/.style={right,at={(-0.0,1.1)}},
title={b)},
legend columns=2,
legend style={at={(0.03,0.03)}, anchor=south west, legend cell align=left, align=left, draw=white!15!black, font=\footnotesize} 
]

\addplot [color=mycolor2, line width=1.0pt]
  table[row sep=crcr]{%
270	1127.94011609211\\
275	1124.42952887082\\
280	1120.92818236383\\
285	1117.43353946207\\
290	1113.94299757413\\
295	1110.45388089963\\
300	1106.96343400392\\
305	1103.46881640518\\
310	1099.96709796056\\
315	1096.45525488038\\
320	1092.93016625158\\
325	1089.38861096663\\
330	1085.82726500681\\
335	1082.2426989636\\
340	1078.63137586783\\
345	1074.98964917091\\
350	1071.31376096603\\
355	1067.59984036031\\
360	1063.84390202139\\
365	1060.04184505623\\
370	1056.18945182532\\
375	1052.28238704794\\
380	1048.31619737409\\
385	1044.28631110719\\
390	1040.18803785063\\
395	1036.0165688465\\
400	1031.76697793624\\
405	1027.43422216575\\
410	1023.01314342167\\
415	1018.49847005235\\
420	1013.88482022397\\
425	1009.16670438224\\
430	1004.33853002842\\
435	999.394602439823\\
440	994.329134716977\\
445	989.136251320456\\
450	983.809995684085\\
455	978.344336162323\\
460	972.73317689134\\
465	966.970365664705\\
470	961.04970057141\\
475	954.964968310054\\
480	948.709904778011\\
485	942.278237244036\\
490	935.663756687476\\
495	928.86019628983\\
500	921.861385675351\\
505	914.661188081705\\
510	907.253537433176\\
515	899.632536013316\\
520	891.792270455668\\
525	883.727090251701\\
530	875.431447708846\\
535	866.899968977429\\
540	858.127508199702\\
545	849.109105792611\\
550	839.840147637516\\
555	830.316180146709\\
560	820.533144778029\\
565	810.487172653006\\
570	800.174792838059\\
575	789.593099827568\\
580	778.739345324598\\
585	767.611487977766\\
590	756.206859757936\\
595	744.524584434482\\
600	732.563689556067\\
605	720.323563753588\\
610	707.803943416564\\
615	695.006081009199\\
620	681.929848221022\\
625	668.57721133838\\
630	654.949879742871\\
635	641.051169333919\\
640	626.883403391181\\
645	612.449780839887\\
650	597.755827923958\\
655	582.805188256601\\
660	567.603773611009\\
665	552.158380628597\\
670	536.475107152007\\
675	520.562304116017\\
680	504.428239638558\\
685	488.084167146817\\
690	471.542715525261\\
695	454.81775022637\\
700	437.928335790681\\
705	420.896951960857\\
710	403.756715703408\\
};
\addlegendentry{\scriptsize fixE}

\addplot [color=mycolor3, line width=1.0pt]
  table[row sep=crcr]{%
270	1128.34093798069\\
275	1124.59472179173\\
280	1120.92526779919\\
285	1117.31590382811\\
290	1113.75233949619\\
295	1110.22225558433\\
300	1106.71497554328\\
305	1103.22120059084\\
310	1099.73279449614\\
315	1096.24260750983\\
320	1092.7443313918\\
325	1089.23237931516\\
330	1085.70178583685\\
335	1082.14812308647\\
340	1078.56743029628\\
345	1074.9561541695\\
350	1071.31109829513\\
355	1067.6293800032\\
360	1063.90839345211\\
365	1060.14577808511\\
370	1056.33939125073\\
375	1052.4872846421\\
380	1048.58768412335\\
385	1044.63897216099\\
390	1040.63967227591\\
395	1036.58843583475\\
400	1032.48403077304\\
405	1028.32533120676\\
410	1024.11130884174\\
415	1019.8410251445\\
420	1015.51362539433\\
425	1011.1283325067\\
430	1006.68444281301\\
435	1002.18131938134\\
440	997.61839168459\\
445	992.995150866031\\
450	988.311147646667\\
455	983.56598914167\\
460	978.759338413668\\
465	973.890912052342\\
470	968.960477102283\\
475	963.967863755303\\
480	958.912938152844\\
485	953.795617511331\\
490	948.615895817611\\
495	943.373780189189\\
500	938.069354578996\\
505	932.70274333015\\
510	927.27412045961\\
515	921.783740636708\\
520	916.231859199132\\
525	910.618833914153\\
530	904.945054799318\\
535	899.21096482314\\
540	893.41707182878\\
545	887.563927427135\\
550	881.652173302279\\
555	875.682474003448\\
560	869.655586977603\\
565	863.57229285065\\
570	857.433455449645\\
575	851.240062759389\\
580	844.993103785711\\
585	838.693721426594\\
590	832.342846786094\\
595	825.941855242609\\
600	819.492046870104\\
605	812.994774801886\\
610	806.451444285629\\
615	799.863800136277\\
620	793.233214904628\\
625	786.561546589682\\
630	779.850566395844\\
635	773.10240453134\\
640	766.318949742417\\
645	759.502296788553\\
650	752.655072076261\\
655	745.779495422578\\
660	738.878323889938\\
665	731.954547962787\\
670	725.011062991454\\
675	718.051329021046\\
680	711.078881486057\\
685	704.098013053921\\
690	697.113452598118\\
695	690.13037149827\\
700	683.15527324379\\
705	676.195674659499\\
710	669.261742146561\\
};
\addlegendentry{\scriptsize Eureqa}

\addplot [color=mycolor4]
  table[row sep=crcr]{%
270	1128.04173090871\\
275	1124.53729779221\\
280	1121.04177803525\\
285	1117.55253746682\\
290	1114.06687825694\\
295	1110.58203147722\\
300	1107.09515101048\\
305	1103.60330851156\\
310	1100.10348919952\\
315	1096.59258830522\\
320	1093.06740805142\\
325	1089.52465505872\\
330	1085.96093812355\\
335	1082.37276625142\\
340	1078.75654701163\\
345	1075.1085850594\\
350	1071.42508091078\\
355	1067.70212988287\\
360	1063.93572122214\\
365	1060.12173757718\\
370	1056.25595442644\\
375	1052.33403981169\\
380	1048.35155455316\\
385	1044.30395263813\\
390	1040.18658155912\\
395	1035.99468336454\\
400	1031.72339635921\\
405	1027.36775648595\\
410	1022.92269976936\\
415	1018.38306478461\\
420	1013.74359690643\\
425	1008.9989517136\\
430	1004.14370076768\\
435	999.172333360356\\
440	994.079267711567\\
445	988.858855683557\\
450	983.50539167087\\
455	978.013119871713\\
460	972.376246620179\\
465	966.588949769746\\
470	960.645386871652\\
475	954.539735852063\\
480	948.266156280529\\
485	941.818834067924\\
490	935.192057607898\\
495	928.380094690159\\
500	921.377352970715\\
505	914.178317358132\\
510	906.777590237307\\
515	899.169994829046\\
520	891.350386841455\\
525	883.313945457737\\
530	875.056010021338\\
535	866.572156260425\\
540	857.858255462856\\
545	848.910433882785\\
550	839.725242052123\\
555	830.299464815117\\
560	820.630369943951\\
565	810.715496624118\\
570	800.552877387002\\
575	790.141218744814\\
580	779.479476462025\\
585	768.567438677264\\
590	757.404323791622\\
595	745.991336810168\\
600	734.329684062283\\
605	722.421060336683\\
610	710.267640221065\\
615	697.873325420617\\
620	685.240676131769\\
625	672.374608773125\\
630	659.279906176638\\
635	645.963211485537\\
640	632.430265121065\\
645	618.687905076088\\
650	604.745637276805\\
655	590.611138977265\\
660	576.294718180628\\
665	561.807851045746\\
670	547.161491447943\\
675	532.369268126606\\
680	517.44498942344\\
685	502.405986634473\\
690	487.27141212138\\
695	472.062124596731\\
700	456.8050518169\\
705	441.531330568304\\
710	426.284302366029\\
};
\addlegendentry{\scriptsize fixE-5}

\addplot [color=mycolor6]
  table[row sep=crcr]{%
270	1128.03415551651\\
275	1124.43816584712\\
280	1120.88027895629\\
285	1117.35284021267\\
290	1113.84883593758\\
295	1110.36177318078\\
300	1106.88557911049\\
305	1103.4145161557\\
310	1099.94310990747\\
315	1096.46608742635\\
320	1092.97832410962\\
325	1089.47479763774\\
330	1085.95054784164\\
335	1082.40064148467\\
340	1078.82014129639\\
345	1075.20407851737\\
350	1071.54742855158\\
355	1067.8450892444\\
360	1064.09186148156\\
365	1060.28243199097\\
370	1056.41135776057\\
375	1052.47305222964\\
380	1048.46177328911\\
385	1044.37161269634\\
390	1040.19648660277\\
395	1035.93012788709\\
400	1031.56608023313\\
405	1027.09769297105\\
410	1022.51811808364\\
415	1017.82030835979\\
420	1012.99701861763\\
425	1008.04080729423\\
430	1002.94404196489\\
435	997.698901669845\\
440	992.297391459656\\
445	986.731351328945\\
450	980.992471695927\\
455	975.072308781911\\
460	968.962308233964\\
465	962.653827303505\\
470	956.13815735221\\
475	949.406592856394\\
480	942.450395789105\\
485	935.260873406576\\
490	927.829509627593\\
495	920.147812242201\\
500	912.207569992662\\
505	904.000791330631\\
510	895.519794230256\\
515	886.757403540805\\
520	877.706688849178\\
525	868.361465502643\\
530	858.716076369288\\
535	848.765555784231\\
540	838.50577471575\\
545	827.933417075988\\
550	817.046323937661\\
555	805.843207627781\\
560	794.324142762506\\
565	782.490231065288\\
570	770.344051996607\\
575	757.890062596514\\
580	745.13384136748\\
585	732.083247254171\\
590	718.745707127865\\
595	705.13326327056\\
600	691.258726703261\\
605	677.13663534075\\
610	662.783239895281\\
615	648.219099132657\\
620	633.462720384972\\
625	618.538209626455\\
630	603.470071667987\\
635	588.287389021595\\
640	573.017876639965\\
645	557.692051014935\\
650	542.346595506983\\
655	527.014493530713\\
660	511.73458288162\\
665	496.548250424877\\
670	481.495517989425\\
675	466.621867672791\\
680	451.972498630391\\
685	437.599420827301\\
690	423.557333129357\\
695	409.902925161799\\
700	396.703929025613\\
705	384.034231661629\\
710	371.990266672034\\
};
\addlegendentry{\scriptsize freeE-5}

\addplot [color=mycolor1, line width=1.0pt]
table[row sep=crcr]{%
	270	1129.6\\
	275	1126\\
	280	1122.5\\
	285	1119\\
	290	1115.5\\
	295	1112\\
	300	1108.5\\
	305	1105\\
	310	1101.5\\
	315	1097.9\\
	320	1094.4\\
	325	1090.8\\
	330	1087.2\\
	335	1083.6\\
	340	1080\\
	345	1076.4\\
	350	1072.7\\
	355	1069.1\\
	360	1065.4\\
	365	1061.6\\
	370	1057.9\\
	375	1054.1\\
	380	1050.3\\
	385	1046.5\\
	390	1042.6\\
	395	1038.7\\
	400	1034.8\\
	405	1030.8\\
	410	1026.8\\
	415	1022.8\\
	420	1018.7\\
	425	1014.6\\
	430	1010.5\\
	435	1006.3\\
	440	1002\\
	445	997.75\\
	450	993.42\\
	455	989.04\\
	460	984.61\\
	465	980.12\\
	470	975.59\\
	475	970.99\\
	480	966.34\\
	485	961.63\\
	490	956.85\\
	495	952.02\\
	500	947.11\\
	505	942.13\\
	510	937.08\\
	515	931.96\\
	520	926.75\\
	525	921.47\\
	530	916.09\\
	535	910.63\\
	540	905.08\\
	545	899.43\\
	550	893.68\\
	555	887.82\\
	560	881.85\\
	565	875.76\\
	570	869.55\\
	575	863.21\\
	580	856.73\\
	585	850.11\\
	590	843.33\\
	595	836.39\\
	600	829.28\\
	605	821.98\\
	610	814.48\\
	615	806.77\\
	620	798.83\\
	625	790.64\\
	630	782.18\\
	635	773.43\\
	640	764.34\\
	645	754.9\\
	650	745.05\\
	655	734.75\\
	660	723.93\\
	665	712.52\\
	670	700.43\\
	675	687.52\\
	680	673.64\\
	685	658.57\\
	690	641.99\\
	695	623.45\\
	700	602.22\\
	705	577.04\\
	710	545.33\\
};
\addlegendentry{\scriptsize REFPROP}

\addplot [color=black,  line width=1.0pt, only marks, mark size=2.8pt, mark=x]
table[row sep=crcr]{%
	719.00	364.96\\
};
\addlegendentry{\scriptsize crit}
\end{axis}
\end{tikzpicture}%

%% file: 05_conclusion.tex
To decrease the time and even financial expenditure for accurate thermodynamic property modeling, it is necessary to gather suitable experimental data with respect to accuracy and information content.
Therefore, we investigated the application of OED to liquid-phase densities of ethylene glycol by calculating the information content of existing experiments along different isotherms.
With our resulting model fitted to the most informative five isotherms, we reproduce the measured data with a maximum relative deviation of \SI[round-mode=figures,round-precision=2]{0.02542}{\%}.
Compared to the maximum relative deviation of the model involving all eight isotherms of \SI[round-mode=figures,round-precision=2]{0.02243}{\%}, this is a very promising result.
These results are validated using the measured data of propylene glycol (\SI[round-mode=figures,round-precision=2]{0.015707}{\%} compared to \SI[round-mode=figures,round-precision=2]{0.014343}{\%}).
The smaller deviations for propylene glycol can be explained by the model development originally based on these data.
By fitting two new models using the best calculated selection of isotherms, we demonstrate that, with OED, sufficiently accurate models can be developed with fewer experiments than traditionally used.

Due to the limitation to already measured data, we are not able to compare these results with a model fitted to the best five freely selected isotherms.
At this point, new measurements become necessary in order to compare the models with the best five measured and the best five freely selected isotherms, where we are confident to achieve even better results.
Finally, the comparison of the fit with free exponents shows a significant improvement compared to the fixed exponent model, which highlights the importance of model selection, even if this model should not be used outside the measured \((p,T)\)-range.
The investigation of the extrapolation behavior underlines this even more, where the use of different thermodynamic criteria beyond the actual measurement data is crucial.

A fortunate circumstance, which should be further investigated, is that the isotherms at ambient temperature are selected last.
From an experimental point of view this is advantageous because these are measurements with larger uncertainty since a thermostating task at ambient temperature is often tricky (\eg, a circulating thermostat has to switch between heating and cooling against the ambient temperature).

Our next steps are:
\begin{itemize}
	\item 
		Measuring the free calculated isotherms, fit the data to the new model and compare the results to the existing models.
	\item
		Defining different objective functions in addition to those describing the parameter uncertainty.
	\item
		Developing methods to include thermodynamic criteria as boundaries for the model (\eg, extrapolation behavior).
	\item
		Using OED based on the nonlinear model (with free exponents) using the sequential process described in \cref{section:introduction}.
	\item
		Investigating the influence of the number of terms to deal with the problem of over-fitting.
\end{itemize}
With these next steps, we aim to create the basis for an OED setup specialized for the measurement of thermodynamic properties.
The data for this paper and the calculations performed are provided in the supporting information.